\documentclass[structabstract,a4paper]{aa}  
\usepackage{lscape}
\usepackage{longtable}
\usepackage{graphicx}
\usepackage{txfonts}
\usepackage{rotating}

 \usepackage{natbib,twoopt}
\usepackage[breaklinks=true]{hyperref} %% to avoid \citeads line fills
\bibpunct{(}{)}{;}{a}{}{,} %% natbib format like A&A and ApJ

\begin{document}

\title{The XMM Deep survey in the CDFS V. Iron K lines from active galactic nuclei in the distant Universe \thanks{Based on observations by \textit{XMM-Newton}, an ESA science mission with instruments and contributions directly funded by ESA member states and NASA.}}

\author{S. Falocco 
  \inst{1}
  \and
  F. J. Carrera
  \inst{1}
  \and
  A. Corral
  \inst{2}
 \and
X. Barcons
\inst{1}
  \and
  A. Comastri
  \inst{3}
  \and
  R. Gilli 
  \inst{3}
  \and
  P. Ranalli
  \inst{2,3}
  \and
  C. Vignali
  \inst{4,3}
\and
 K. Iwasawa
  \inst{5}
 \and
  N. Cappelluti
\inst{3}
 \and
 E. Rovilos
  \inst{6}
\and
I. Georgantopoulos 
 \inst{2,3}
 \and
M. Brusa 
\inst{4,3,7}
 \and
F. Vito
\inst{4,3}  
}
\institute{Instituto de F\'isica de Cantabria (CSIC-UC)
   39005 Santander, Spain \\
  \email{falocco@ifca.unican.es}
\and
Institute of Astronomy \& Astrophysics, National Observatory of Athens, Palaia Penteli, 15236, Athens, Greece
  \and
INAF-Osservatorio Astronomico di Bologna, via Ranzani 1, 40127, Bologna, Italy
\and
Dipartimento di Fisica ed Astronomia, Universit\`a di Bologna, Via Berti Pichat 6/2, 40127 Bologna
\and
 ICREA and Institut de Ci\`encies del Cosmos (ICC), Universitat de Barcelona (IEEC-UB),  08028, Barcelona, Spain
\and
Department of Physics, University of Durham, South Road, Durham DH1 3LE, UK
\and
Max Planck Institute f\"ur Extraterrestriche Physik. 85748 Garching, Germany
}

\date{Received  2013-01-11; accepted 2013-04-22}

% \abstract{}{}{}{}{} 
% 5 {} token are mandatory

\abstract
{X-ray spectroscopy of active galactic nuclei (AGN) offers the opportunity to directly probe the inner regions of the accretion disk.
 Reflection of the primary continuum on the circumnuclear accreting matter produces features in the X-ray spectrum that help to explore the physics and the geometry of the innermost region, close to the central black hole.}
{We present the results of our analysis of average AGN \textsl{XMM-Newton} X-ray spectra in the \textsl{Chandra} Deep Field South observation (hereafter, XMM CDFS), in order to explore the Fe line features in distant AGN up to z$\sim$3.5. 
%Our main goal is to constrain the iron line properties over several orders of magnitude in the 2-10 keV X-ray luminosity band (10$^{41}$ to 10$^{45}$ erg $\rm s^{-1}$) and a broad range of redshift (up to 3.6). For that purpose we use a total of 100 high quality X-ray spectra of a sample of 51 distinct sources.
}
{We computed the average X-ray spectrum of a sample of 54 AGN with spectroscopic redshifts and signal-to-noise ratio in the 2-12 keV rest-frame band larger than 15 in at least one EPIC camera (for a total of 100 X-ray spectra and 181623 net counts in the 2-12 keV rest-frame band).
We have taken the effects of combining spectra from sources at different redshifts and from both EPIC-pn and EPIC-MOS cameras into account, as well as their spectral resolution; we checked our results using thorough simulations.
%We also developed a new model-independent method to estimate their equivalent width (EW).
We explored the iron line components of distant AGN focusing on the narrow core which arises from material far from the central BH and on the putative relativistic component emitted in the accretion disk.
}
{The average spectrum shows a highly significant iron feature. Estimating its equivalent width (EW) with a model-independent method suggests a higher EW in a broader range. The line, modelled as an unresolved Gaussian, is significant at 6.8$\sigma$ and has an EW=$95\pm22$~eV for the full sample. We find that our current data can be fitted equally well { adding a relativistic profile to the narrow component (in the full sample, EW=$140\pm120$~eV and 67$\pm28$ eV respectively for the relativistic and narrow lines)}.}
% conclusions heading (optional) leave it empty if necessary
{Thanks to the high quality of the XMM CDFS spectra and to the detailed modelling of the continuum and instrumental effects, we have shown that the most distant AGN exhibit a highly significant iron emission feature.
It can be modelled both with narrow and broad lines which suggest that the EW becomes higher when a broader energy range around the line centroid is considered, provides tantalising evidence for reflection by material both very close and far away from the central engine. The EW of both features are similar to those observed in individual nearby AGN, hence they must be a widespread characteristic of AGN, since otherwise the average values would be smaller than observed. }

\titlerunning{Iron lines in the XMM CDFS deep survey}
\maketitle
    %
    %________________________________________________________________
   
\section{Introduction}

X-ray spectroscopy of active galactic nuclei (AGN) provides
one of the most useful probes of the properties of the accretion disks
in the vicinity of the super massive black holes (SMBHs). 
In particular, the iron K$_{\alpha}$ line is widely
used to study the SMBH relativistic effects, therefore
giving unique information about the location and the kinematics of the
accreting material.  Relativistic effects produce a line deformation when 
 fluorescence occurs in regions a few gravitational radii away from the central SMBH. Given that the inner radius of the accretion disk is 6$R_g$ around a non-rotating black hole (BH)
 and 1.23 $R_g$ ($R_g$=$\frac{GM}{c^2}$) if the BH is maximally rotating
\citep{bardeen}, the velocities involved approach the speed of light and, furthermore, the strong gravitational field of the BH itself will cause a gravitational redshift effect.
Since the theoretical predictions on AGN relativistic lines \citep{fabian1989}, such features have been found in the X-ray spectra of several Seyferts, being MGC-6-30-15 the best studied one \citep{tanaka}.
%%  The first AGN with a relativistic broad Fe line to be discovered was MGC 6-30-15, studied multiple
%% times by the scientific community. Already in its first
%% observations made by \textsl{GINGA} \citep{pounds1990} and \textsl{ASCA},
%% \citep{tanaka}, a broad and intense iron line
%% was detected. Later observations of \textsl{XMM-Newton} confirmed that the shape of the line was due to relativistic effects. The
%% presence of a relativistic line, as described in 
%% %\cite{reynolds} and
%% \cite{wilms2001} and the inner radius measured in X-rays probed the
%% presence of a rapidly spinning SMBH \citep{reynolds2004}. After the first results on this
%% Seyfert nucleus, other low redshift AGN observed with high Signal-to-Noise Ratio (SNR)
%% showed relativistic Fe line profiles. In the high redshift Universe, the detection of broad lines is more challenging given the limitations of X-ray instruments. However, an interesting case has been found in \cite{comastri2004}: a broad iron line was detected in an AGN spectrum at redshift $\sim$1.15 in the \textsl{Chandra} Deep Field North Survey.

A broad Fe K$\alpha$ line can be diluted in the continuum if the signal-to-noise ratio (S/N) is low \citep{guainazzi2006}. This demonstrates the necessity of high S/N data with good counts statistics to detect the relativistic features in the iron lines, \citep{guainazzi2011,delacalle2010,nandra2007}. 
Theoretical predictions \citep{ballantyne2010} supported this scenario, showing that the majority of relativistic iron lines have equivalent widths (EW) of just 100 eV, highlighting the importance of high sensitivity surveys to explore them. 
In cases where
the individual X-ray observations have a limited S/N, statistical methods like
integrating or averaging the spectra can be used for their
analysis. \cite{nandra97} used for the first time a statistical
approach to study the \textsl{ASCA} X-ray spectra of an AGN survey: they found that a
broad line is present not only in the majority of their Seyfert sample, but also in their average spectrum, with a shape similar to those of the individual source spectra
probing therefore that  these features are common in the low redshift universe. Their more recent work \citep{nandra2006} on an \textsl{XMM-Newton} broad line survey confirmed this result, finding that in 1/3 of their sample the iron lines are well described by the sum of a narrow line likely coming from reflection in distant cold matter in the inner edge of the torus, and a relativistic line coming from the accretion disk, much closer to the SMBH.

The deep observation with \textsl{XMM-Newton} of the Lockman Hole was studied
with a statistical approach, showing a broad stacked iron line, with a
relativistic shape \citep{alina}.  Some further tantalising evidence for a broad iron line was also found in
\cite{brusa2005} and \cite{civano}: they computed the stacked
X-ray spectra of the \textsl{Chandra} Deep Fields, cautioning that
their large EW and broad profile significantly depend on the modelling
of the underlying continuum.
In fact, the EW of the lines found in these early works were much higher (of the order of many 100 eV) than those found in local individual AGN, which casted some doubts on their reality.

Unlike these first spectroscopic studies of the AGN surveys with stacking, many recent papers have successfully detected narrow iron lines, while the broad lines are not clearly seen. Examples of such detections include the \textsl{XMM-Newton} X-ray observations of the AXIS \citep{carrera2007} and XWAS samples studied by \cite{corral2008} and the subsample with 0.2-12 keV counts greater than 1000 of the 2XMM X-ray source catalogue \citep{watson2008} studied by \cite{chaudhary2010,chaudhary2012}. Therefore, the detection of broad Fe lines in stacked spectra of distant AGN remains controversial.

More recently, \cite{iwasawa2011} found evidence for ionised iron lines in the stacked
spectrum of the COSMOS sample. 
Broad symmetrical lines were found in the low luminosity - low redshift AGN in the deep \textsl{Chandra} Fields \citep{falocco2012}.
Taking the COSMOS results into account, such symmetrically broadened lines were interpreted as the combination of a relativistic profile and several narrow lines centred at 6.4 keV and higher energies, from iron at a variety of ionisation states. Consequently the total iron line emission will appear symmetrically smoothed \citep{falocco2012}.\\
In this paper, we present the result of an X-ray spectral analysis of the deepest \textsl{XMM-Newton} survey made so far: the 3 Ms observation centred in the \textsl{Chandra} Deep Field South \citep{comastri11}. 
This survey enables a much more detailed spectroscopic study than earlier data on this or other deep surveys and allows us to study the accretion at the epoch of the peak of the activity in the history of galaxy evolution at z$\sim$1-2 \citep{ueda,ebrero2009}.
It covers three orders of magnitude in luminosity ($10^{41.5}-10^{44.5}\rm erg$ $\rm s^{-1}$), which allows us to study the behaviour of the spectral features coming from the accretion disk in X-rays, and the trend found between the Fe line EW with the luminosity \citep{iwasawa94}.

The paper is organised as follows. The properties of the sample are
described in Sect. \ref{survey}, the method used to average the spectra in Sect. \ref{method}; our results are illustrated in Sect. \ref{results}; and in Sect. \ref{discussion} we discuss the implications, summarising our conclusions in Sect. \ref{conclusions}.
Throughout this paper, we adopt the
cosmological parameters: $H_{\ 0}$=70 km s$ ^{-1}$ Mpc$^{-1}$, $\Omega_{matter}$=0.3 and 
$\Omega_{\Lambda}$=0.7; all the counts
refer to the net number of counts between 2 and 12 keV rest-frame, unless explicitly stated otherwise. 
We used \texttt{xspec v. 12.5}
\citep{arnaud} for the X-ray spectral analysis; the error estimates correspond to 90\% confidence level for one interesting parameter. 

%________________________________________________________________
\section{X-ray sample}\label{survey}

We present in this section the \textsl{XMM-Newton} observation in the CDFS and its spectral analysis in \ref{parent_survey}, the definition of the 54 sources sample in \ref{def_survey}, and the definition of the various subsamples in \ref{sub_survey}.

\begin{table*}\label{tpropertiesS/N15}

\centering
  \caption{Properties of the full sample with S/N$>15$ and its subsamples}
%\small\centering
\centering
\begin{tabular}{lrrrrrrrrrrr}
%\hline\hline\noalign{\smallskip}
Sample &  $N_{sp}$ & N$_s$ &   $N_{2-12}$  & $N_{5-8}$   & $\langle z \rangle$ & $\langle \log(L) \rangle $ & $ \langle N_{\rm H,22} \rangle $  & $ N_{\rm H,22} $  & $\langle S/N \rangle$ & $\alpha$ & $\Sigma_6$ \\  
\hline
 & & & &    &  &  (erg s$^{-1}$)  & ($10^{22}$ cm$^{-2}$)  & ($10^{22}$ cm$^{-2}$)    & & &  (eV) \\  
 (1) & (2) & (3) & (4)   & (5) &  (6)  & (7) & (8) & (9)  & (10) & (11) & (12) \\
\hline
Full   & 100 &  54 &  181623  & 40863  & 1.34  & 43.74   & 1.48  &  0.08  &   29.8  &  0.35 &  119  \\
  \hline
%$\log{\left( N_{\rm H}/{\rm cm^2}\right)}<21.5$
Unabs & {52} & {32} & 110523 & 21071 & 1.27 & 43.78 & 0.02 & 0.0 & 33.60 & 0.34 & 122\\
%Unabs & 68 & 38 & 131383  & 25532  & 1.37   &  43.78   &  0.05 & 0.  & 31.5 & 0.34  & 121  \\
%  $\log{\left( N_{\rm H}/{\rm cm^2}\right)}\geq21.5$ 
Abs   & {29} & {17} &  45896 & 14373 & 1.28 & 43.64 & 4.90 & 2.41 & 24.88 & 0.37 & 116 \\
%Abs   & 32 & 16  & 50240  & 15331  & 1.30   &  43.64   & 4.51 & 2.20 &  24.6 & 0.37  &  116 \\
 \hline
z$<1$ &39 & 20 & 72672  & 18148 & 0.64 &  43.21 & 1.61  & 0.17  & 30.0   & 0.35 &   95\\
z$\geq1$ &61 & 34 &  108951   &  22715 & 1.80 & 44.07  & 1.39    & 0.04   & 28.8 & 0.35 &  134 \\

 \hline
%$\log{\left( L/{\rm erg~s^{-1}}\right)}<44$
Sey & 61 & 33 & 91511 & 20987 &  0.91 & 43.35 & 1.35 & 0.11   & 26.2 &  0.34 & 108  \\
%$\log{\left( L/{\rm erg~s^{-1}}\right)}\geq44$
QSO  & 39 & 21  & 90111  & 19876 & 2.02 & 44.34 &  1.68 & 0.04  &  34.1 & 0.36  & 138 \\
%%    \hline
%% z$<=0.837$      & 35  & 60329  & 14689  &  0.599  & 43.10    &  1.33   & 28.00 & 0.35  & 93.0  \\
  
%% $0.976<$z$<1.605$     & 33  & 57473  & 12444  & 1.24   &  43.847   &     1.89   & 28.66  & 0.36  & 111.3 \\

%% $1.608<$z$<3.61$       & 32  & 63820  & 13730  & 2.26   &  44.32   &   1.20   &  31.29 & 0.38  &  139.1\\
%%    \hline
%%  $41.53<$log(L)$\leq43.70$        & 45  & 59234  & 14121  & 0.78   & 43.167    &  0.81   & 23.24   & 0.34  & 99.0  \\
 
%% $43.72<$log(L)$<44.22$      & 31  & 59392  & 11930  & 1.54   &  43.98   &  1.82   & 31.40   & 0.35  &  116.7\\

%% log(L)$>44.23$            & 24   & 62997  & 14812  & 2.14   &  44.49   &  2.27   &  37.85  & 0.36  &  129.4\\

\hline
\hline
\end{tabular}

\tablefoot{Unabs and Abs subsamples have and $\log{\left( N_{\rm H}\right)}<21.5$ and $\log{\left( N_{\rm H}\right)}\geq21.5$ respectively; Sey and QSO subsample have $\log{\left( L\right)}<44$ and $\log{\left( L\right)}\geq44$ respectively (see text). Columns: (1) (Sub)Sample; (2) Number of spectra; (3) Number of sources; (4) number of net counts in 2-12 keV rest-frame; (5) number of net counts in 5-8 keV rest-frame; (6) average redshift; (7) Logarithm of the average  rest-frame 2-10~keV luminosity in units of erg s$^{-1}$, corrected for Galactic and intrinsic absorption; (8) and (9) average and median of the intrinsic column density in $10^{22}$ $\mathrm{cm^{-2}}$; (10) average S/N of the (sub)sample between 2 and 12 keV rest-frame; (11) slope of the power-law dependence between the width of the simulated unresolved lines and their central energies $\Sigma=E^{-\alpha}$ (see Sect. 3.3); (12) width at 6 keV of the simulated iron lines (see Sect. 3.3)}
\end{table*}

\subsection{The X-ray data}\label{parent_survey}

The XMM CDFS survey is an observation of the \textsl{Chandra} Deep Field South with \textsl{XMM-Newton} for a total of 3.3 Ms of full exposure time. Exposure times from the EPIC pn and MOS cameras are 2.5 Ms and 2.8 Ms respectively. The bulk of the observations were
performed between July 2008 and March 2010, and have been combined
with archival data taken between July 2001 and January 2002. The
details on data reduction and analysis and the source detection algorithm are presented in \cite{ranalli}. 

 In this work, we used the high quality X-ray spectra for
sources detected with a significance above 8 sigma in the 2-10 keV source catalogue, corresponding to a flux
limit of $\sim$ 2 $\times$ $10^{-15}$erg s$^{-1}$cm$^{-2}$.

 We looked for spectroscopic identifications of
 these sources in the literature making use of several spectroscopic
 campaigns 
\citep{z_lefevre2,z_szokoly,z_lefevre,z_mignoli,z_ravikumar,z_kriek,z_vanzella,z_taylor,z_treister,z_vandervel,z_balestra,z_silverman,z_casey,z_cooper}.
 To avoid possible spurious broad line detections caused by errors in
 the redshift estimate, which could produce a false line broadening,
 we used only sources with spectroscopic redshifts defined as secure in the literature.
% The \textsl{Chandra} X-ray positions were used to find the correct optical-infrared counterpart, using the likelihood ratio method by Ranalli et al. (2012, in preparation) and \cite{manolis2012}.

To characterise the spectra and to estimate their intrinsic luminosities, after grouping the \textsl{XMM-Newton} EPIC MOS and PN spectra at 20 counts per bin, the spectra were fitted with \texttt{xspec} using the $\chi^{2}$
statistics
between 1 and 12 keV rest-frame
 in order to fit the broadest possible band avoiding the background-dominated E$>$12 keV rest-frame band (for this spectral region our grouping of 20 counts per bin would not probably be a suitable choice due to the background); we do not consider the soft X-ray band (E$<$1keV) due to the onset of complex spectral features (warm absorption and soft excess). We used a simple powerlaw modified by Galactic absorption, frozen at a column
density $\sim8~\times~10^{19}~$cm$^{-2}$ \citep{stark1992}, and variable intrinsic absorption. We analysed the background spectrum and found an emission feature coincident with the position of the Al line characterising many
background spectra at observed-frame energy $\sim$ 1.5 keV \citep{carter2007}. However, this feature is not expected to influence significantly our iron line analysis, as there is only one object at z higher than 3.27, value which would shift the Al feature to 6.4 keV.
To assess whether this systematic
feature influences the output of our fits, we repeated the fits
removing the channels corresponding to $1.5 \pm 0.1$ keV. 
The fit results show
that the presence of this background feature does not really influence
the output fit parameters very much: comparing for example the intrinsic absorption estimated in X-rays, we found that the difference is negligible in the majority of the sample, with the exception of a handful of
spectra (seven, for which we used the fit parameters obtained ignoring these channels).
The luminosities were estimated from the best
 fit models of each spectrum, corrected for Galactic and intrinsic
 absorption, and calculated between 2 and 10 keV rest-frame. 
These are the luminosities used in the remainder of this paper for each spectrum. In the case of sources with both EPIC MOS and pn spectra, when a single value per source is presented (e.g. Fig. \ref{fig:l_z_chandra}), we have used the value corresponding to the spectrum with the highest S/N.

In the remainder of the paper, the luminosities are always indicated in ${\rm erg}\, {\rm s}^{-1}$  and the intrinsic absorption will always be in units of cm$^{-2}$.

\subsection{Sample definition and properties}\label{def_survey}

It has been found empirically that a good S/N is
essential for a significant detection of a broad line with the EPIC
cameras \citep{delacalle2010,guainazzi2011}. 

  \begin{figure}
\centering
\includegraphics[width=8cm,angle=0]{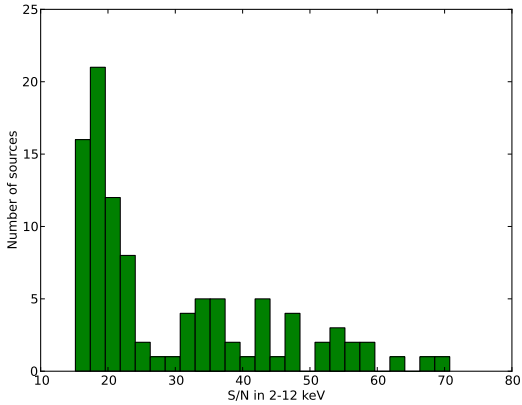}
\caption{Spectral signal-to-noise-ratio between 2-12 keV of the full (100 spectra) sample.}\label{fig:snr}
  \end{figure}

  \begin{figure}
\centering
\includegraphics[width=8cm,angle=0]{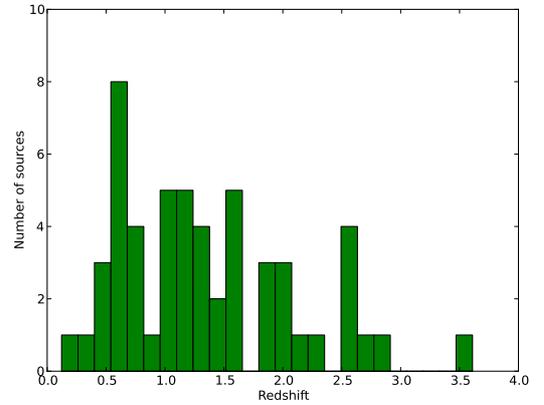}
\caption{Redshift distribution of the full (54 X-ray sources) sample.}\label{fig:z}
  \end{figure}

  \begin{figure}
\centering
  \includegraphics[width=8cm,angle=0]{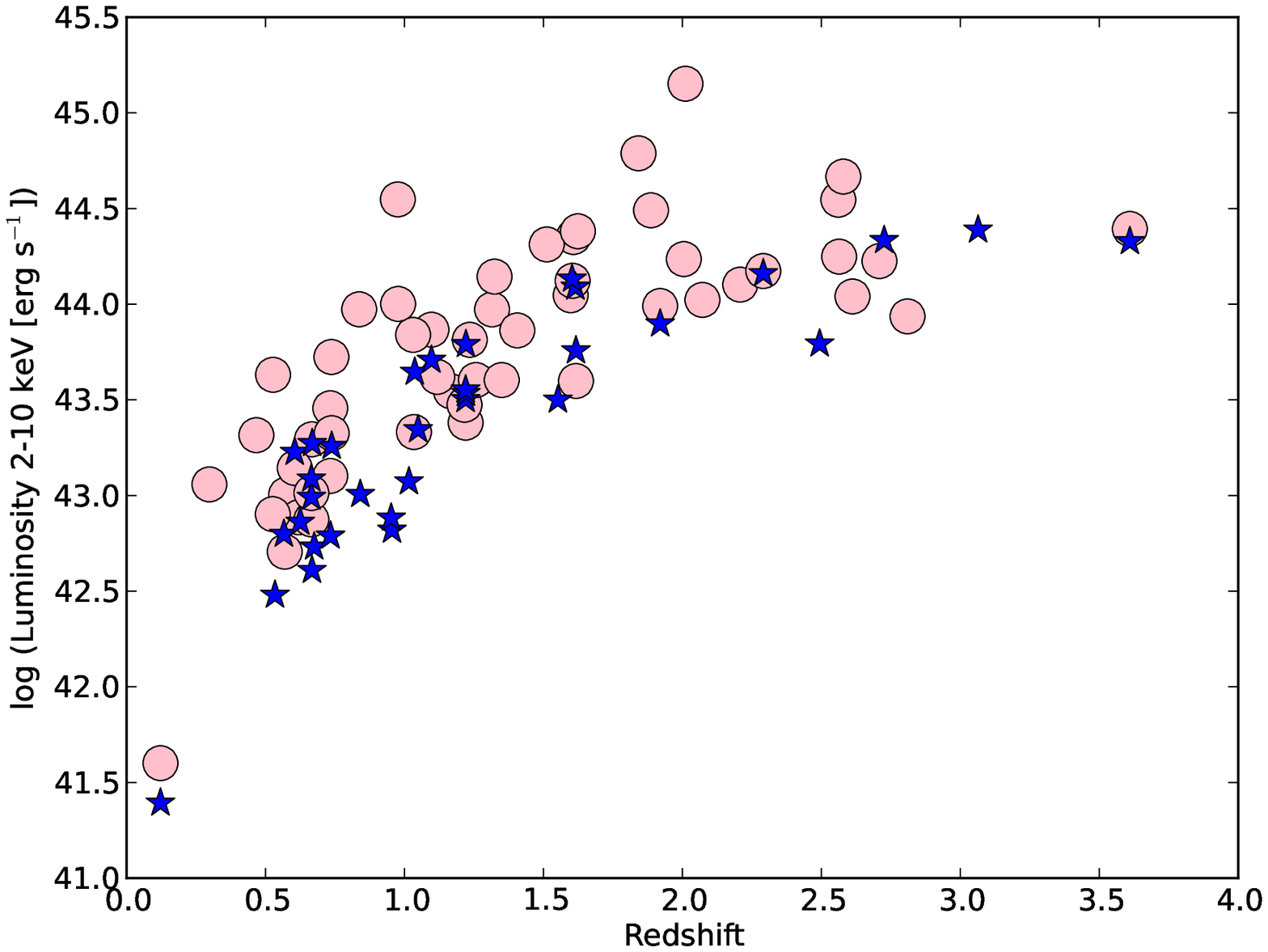}\\
\includegraphics[width=8cm,angle=0]{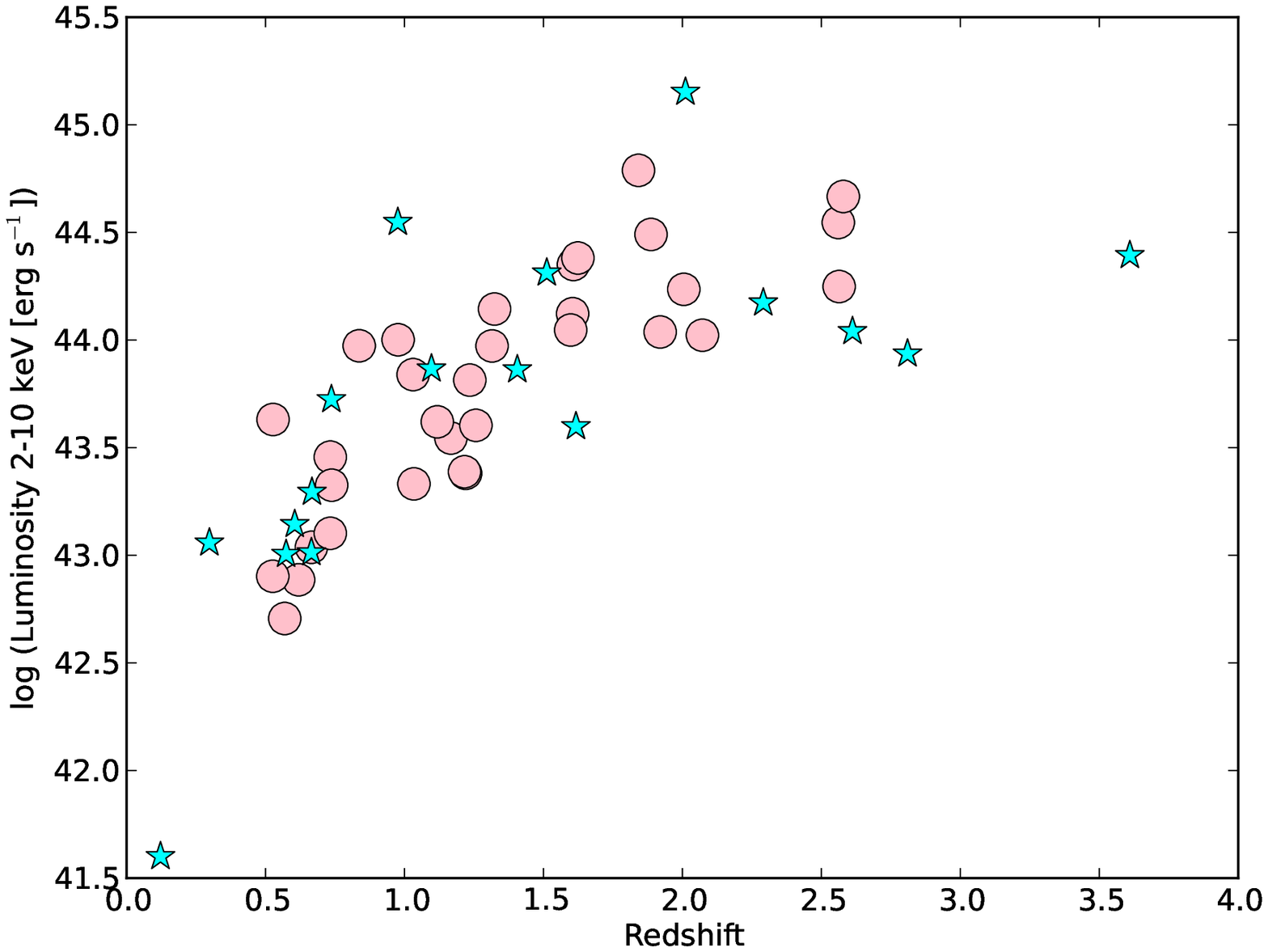}
\caption{Top panel: Distribution in the Luminosity-redshift plane of the 54 XMM CDFS sources (pink circles) and of the 33 Chandra Deep Field South sources (blue stars) by \cite{falocco2012}. Bottom panel: Distribution in the Luminosity-redshift plane of the {17} absorbed (cyan stars) and {32} unabsorbed (pink circles) subsamples. \label{fig:l_z_chandra}}
  \end{figure}

  \begin{figure}
\centering
   \includegraphics[width=8cm,angle=0]{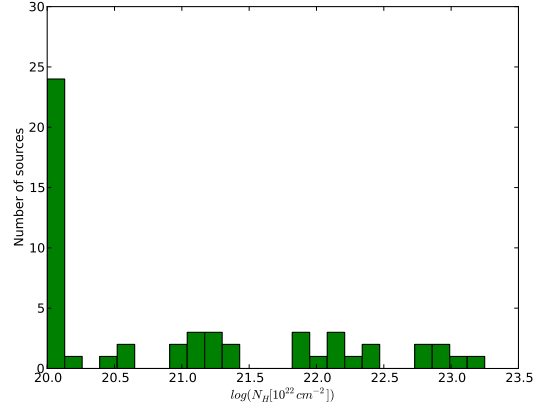}
\caption{$\log{\left( N_{\rm H}\right)}$ distribution of the full sample. $N_{\rm H}$ is in units of $10^{22}$ cm$^{-2}$; all the values $\log{\left( N_{\rm H}\right)} < 20$ have been replaced with $\log (N_{\rm H}) = 20$ for plotting purposes only.}\label{fig:nh}
  \end{figure}

 To maximise the
average S/N of the sample, taking
the level of the background of the \textsl{XMM-Newton} observation into account, we
limited our investigation to the 100 spectra corresponding to 54 unique sources (49 spectra from the EPIC-MOS cameras and 51 spectra from the EPIC-pn camera) characterised with S/N$>$15 estimated between 2 and 12 keV rest-frame (the band used in the analysis of the average spectra described below). 
 This selection of the sample represents a compromise between the need to avoid including noisy sources and the necessity to have a high enough number of sources in the sample for good statistics. We discarded the two brightest sources whose number of counts (about 20000 net counts each) is significantly higher than those characterising the other sources (each of the sources analysed here have less than 10000 net counts in the 2-12 keV rest-frame band and the total accumulated counts by the 54 sources is about 180000). In this way we avoid the estimated average spectra being dominated by the two brightest sources, which would be contributing 10\% of the total counts each. These sources are: PID 319 and PID 203 \citep{ranalli}, the latter shows a remarkable and variable iron line. 
%We included the AGN source PID 146 with $\log{\left( L\right)}$=41.60, to be consistent with} \cite{falocco2012}.
%(PID 319 with RA=53.112 and DEC=-27.685) and (PID 203 RA=53.036 and DEC=-27.793).

The high quality of the spectra can be seen in the Fig. \ref{fig:snr}, where the distribution of the S/N (in 2-12 keV) of the 100 selected spectra is shown. 
We can see in Fig. \ref{fig:z} the distribution of the spectroscopic redshifts of the sources
included in the sample: 21 AGN in our sample are
located at redshifts z$\sim$1-2 and a dozen AGN at earlier epochs ($z\sim 2-3$).
Note, from this distribution, that the rest-frame 2-12 keV band used in this paper, corresponds to an observed band of 1-6 keV at a typical z$\sim$1, therefore avoiding strong fluorescent signatures in the detectors' background.

%%   \begin{figure}\label{fig:counts}
%%    \includegraphics[width=7cm,angle=0]{figures/ra_dec_chandra_xmm_CDFS.ps}
%% \caption{Spatial limits of the sample used in this work (red circles) and of the Chandra Deep Field South spectra used in \cite{falocco2012}}
%%   \end{figure}

%%   \begin{figure}\label{fig:counts}
%%    \includegraphics[width=7cm,angle=0]{figures/nh_cut_nocut.ps}
%% \caption{Logarithmic local absorption estimated in X-rays (see text) ignoring the channels of the Al background spike, in the 'y' axis, and including that channels, in the 'x' axis}
%%   \end{figure}

 %% The ultra-deep exposure and the \texttt{XMM-Newton}
%% sensitivity have allowed to accumulate a high number (100) of good
%% quality spectra (S/N$>15$) in the energy range between 2 and 12 keV.

Figure \ref{fig:l_z_chandra} shows the distribution of the sample sources in the luminosity - redshift plane. In this figure, only one entry per physical source is presented: in sources where both the spectra from MOS and from pn are available, we display the result for the highest S/N spectrum. 
The CDFS observation by \textsl{XMM-Newton} allows a better coverage of the luminosity-redshift plane than our previous work on the \textsl{Chandra} spectra of the sources detected in the deep \textsl{Chandra} fields (see top panel of Fig. \ref{fig:l_z_chandra}). 
The XMM CDFS survey is able to characterise with the highest significance
reached so far the properties of distant X-ray selected AGN over three orders of magnitude
of continuum luminosity and a broad span of redshifts, up to
$\sim3.6$.
Despite the good counting statistics of the individual spectra, the
kind of analysis of the iron line we carry out in this paper requires
much higher statistics, hence we perform stacking. 

%%   \begin{figure}\label{fig:counts}
%%    \includegraphics[width=7cm,angle=0]{figures/log_counts.ps}
%% \caption{Logarithm of the counts between 2-12 keV of our full sample}
%%   \end{figure}

%%   \begin{figure}\label{fig:z}
%%    \includegraphics[width=7cm,angle=0]{figures/L.ps}
%% \caption{Luminosity distribution of our full sample}
%%   \end{figure}

%%   \begin{figure}\label{fig:L}
%%    \includegraphics[width=7cm,angle=0]{figures/log_lum.ps}
%% \caption{Logarithm of the luminosities betwen 2 and 10 keV of our full sample}
%%   \end{figure}

%%   \begin{figure}\label{fig:nh}
%% \centering
%%    \includegraphics[width=10cm,angle=0]{figures/S/N_tot_histo.ps}
%% \caption{S/N of the total sample and its other definition with the condition S/N$>15$}
%%   \end{figure}

%%   \begin{figure}\label{fig:nh}
%%    \includegraphics[width=8cm,angle=0]{figures/S/N_unabs_histo.ps}
%% \caption{S/N of the unabsorbed sample and its other definition with the condition S/N$>15$}
%%   \end{figure}

\subsection{Subsample definition}\label{sub_survey}
The sample characteristics and properties are listed in Table 1. 
 We
 constructed two luminosity subsamples with similar number of counts in
 each, resulting in a Seyfert subsample with $\log{\left( L\right)}<44$ (hereafter, we call refer to this subsample as Sey) and a quasi stellar object (QSO) subsample with $\log{\left( L\right)}\geq44$ (hereafter, we call this subsample QSO).

 We also segregated the full sample in two redshift subsamples with the same number of counts. The low-z subsample includes all the sources with z $<$ 1.
 The high-z subsample includes all the sources with z$\geq1$. 
%We set the threshold at z=1 because the peack of accretion activity has been found for higher redshifts; 
We did not define more than two subsamples because we preferred to have 
as many source counts in each to attempt a sensitive assessment of the iron line trend with these parameters.

Finally, we separated the full sample in two subsamples of different
intrinsic column density estimated in X-rays ('$N_{\rm H}$
subsamples') setting the threshold at $\log{\left( N_{\rm
    H}\right)}=21.5$. This is often used as the threshold below which
the interstellar medium of the galaxies hosting the AGN could provide
the obscuration of broad lines in optical spectra
\citep{caccianiga2007}.  The $N_{\rm H}$ distribution (see
Fig. \ref{fig:nh}) extends from $\log{\left( N_{\rm H}\right)} < 20 $
to $\log{\left( N_{\rm H}\right)} \sim 23$. The $\rm N_H$ shown in the
figure have been obtained from the fits to the individual spectra. For
sources with both MOS and pn spectrum, we present the result from the
best S/N spectrum.  We can also appreciate in this Figure that around
$\log{\left( N_{\rm H}\right)} \sim 21.5$ there is a clear gap between
lower and higher absorption sources, making this value a `natural'
choice for our sample.

{The two $N_{\rm H}$ subsamples have been constructed
  considering the errors} $\Delta N_{\rm H}$ {on the} $N_{\rm
  H}$ {obtained in the fits to the individual spectra. The
  sample, which includes the sources for which at 90 \% of confidence} $\log{\left( N_{\rm H}+\Delta N_{\rm H}\right)}<21.5${ (hereafter called 'Unabs' subsample) accumulates about
  111000 net counts in the 2-12 keV rest-frame band, while the sample
  with }$\log{\left( N_{\rm H}-\Delta N_{\rm H}\right)}\geq21.5${ at 90 \% of confidence
  (called hereafter 'Abs' subsample) contains about 46000 net counts
  in the same band. Nineteen spectra have not been included in the two above
  mentioned subsamples (i.e., the 90\% error bars around the best fit $N_{\rm H}$ values cross the} $\log{\left( N_{\rm H}\right)}=21.5${ borderline that separates 'Abs' from 'Unabs' sources). These unclassified spectra contain 25204 and 5419
  counts respectively in the 2-12 keV and the 5-8 keV rest-frame
  bands. }

We note that, despite its name, the 'Abs' subsample consists mostly of moderately absorbed AGN (median $N_{\rm H}\sim2.4\times10^{22}\rm cm^{-2}$).
Our sample does not contain any Compton thick sources, so we do not expect to detect in the average spectra of the absorbed sample any strong absorption feature in the iron line region.
The two $N_{\rm H}$ subsamples are characterised by a similar distribution in the luminosity-redshift plane (see bottom panel of Fig. \ref{fig:l_z_chandra}), allowing to assess whether the intrinsic absorption alone can influence the detection of the X-ray spectral features of the AGN. In the Fig. \ref{fig:l_z_chandra}, one entry per each physical source is shown (if the spectra from both MOS and pn are available, we plot the best S/N spectrum). In the Fig. \ref{fig:l_z_chandra}, the 54 XMM CDFS sources have higher luminosities at the same redshift range than the 33 \textsl{Chandra} CDFS AGN. According to the Iwasawa-Taniguchi effect, a higher luminosity corresponds to a lower iron line EW, thus the spectra from the XMM CDFS will not be particularly advantaged in the iron feature detection.

%%   \begin{figure}\label{fig:z}
%%    \includegraphics[width=8cm,angle=0]{figures/nh_lum_bins.ps}
%% \caption{NH distribution of the luminosity subsamples}
%%   \end{figure}

%%   \begin{figure}\label{fig:z}
%%    \includegraphics[width=8cm,angle=0]{figures/nh_z_bins.ps}
%% \caption{NH distribution of the luminosity subsamples}
%%   \end{figure}

%%   \begin{figure}\label{fig:z}
%%    \includegraphics[width=8cm,angle=0]{figures/mid_high_map.ps}
%% \caption{Positions in the sky of the middle and high L-z subsamples}
%%   \end{figure}

%%   \begin{figure}\label{fig:z}
%%    \includegraphics[width=8cm,angle=0]{figures/z_lum_bins.ps}
%% \caption{Redshift distribution of the luminosity subsamples}
%%   \end{figure}

%%   \begin{figure}\label{fig:z}
%%    \includegraphics[width=8cm,angle=0]{figures/z_high_bins.ps}
%% \caption{Redshift distribution of the luminosity subsamples}
%%   \end{figure}

%% %____________________
\section{Analysis method}\label{method}

The averaging method used here was presented for the first time \citep{corral2008,corral2011} in the study of \textsl{XMM-Newton} spectra of the XMS-XWAS
surveys \citep{carrera2007,mateos2008} and for the XBS survey \citep{dellaceca2004}; later, it was adapted and tested for the deep \textsl{Chandra}
Fields by \cite{falocco2012}. In this work we refined the technique of analysis introducing new methodologies that will be described in Sect. \ref{simu_line} and \ref{model_ind}.

\subsection{Averaging method}

After fitting the spectra of every individual source as described in Sect. \ref{parent_survey}, we read into
\texttt{xspec} again each unbinned, background-subtracted spectrum,
and we used the corresponding best-fit (to the binned spectrum) model to extract the unfolded spectrum compensating for the effective area of the instrument (\texttt{eufspec} command in
\texttt{xspec}). 
The unfolded spectra depend on the model used for this operation. In \cite{falocco2012}, we have checked the effect of the correction for detector response on the average spectrum, demonstrating the independence on the model used in the unfolding process above $\sim$2 keV, even for models clearly different from the input spectrum. Our simulations (as explained in Sects. \ref{simu_cont} and \ref{simu_line}) take any effect of the uncertainty in the continuum shape into account, since each simulated spectrum is fitted and the corresponding best fit model is used for the unfolding, instead of the input shape.

%%  In this step it is possible that some distortions to narrow spectral features were introduced, such as to the
%% Fe emission line. To quantify this effect, we performed extensive simulations of
%% the continuum and of the Fe Line as described in Sects. 3.2 and 3.3.
Once we obtained all the unfolded spectra, in analogy to what was done in \cite{falocco2012} and in \cite{corral2008}, we applied the following procedure:
\begin{enumerate}
      \item Correction for Galactic absorption,

      \item Shift the spectra to rest-frame,

      \item Re-normalise each spectrum by dividing by its integrated flux in the 2-5, 8-10 keV spectral range
\end{enumerate}

%% In the normalisation process we considered that a correct determination of the continuum excluding any contribution from the putative iron line, can be made integrating the spectra between 2 and 5 and 8 and 10 keV. 

%% For this reason, we computed the sum of the the flux between 2 and 5 keV and the flux  between 8 and 10 keV and finally divided each bin for this value. With the higher number of absorbed sources in the survey, compared with that of \cite{falocco2012}, normalising for the continuum in 2-5 keV would assign a too high weight to the absorbed sources. In this way, the continuum reproduced best the observed spectrum blue-wards of 7 keV.

We used the 2-10 keV rest-frame band in the last step (excluding the 5-8 keV rest-frame range were the putative iron line is expected to contribute most) to avoid giving too much weight to absorbed sources, which are in much higher proportion of the full sample with respect to our previous work in \cite{falocco2012}, where we used only the 2-5 keV rest-frame band for the normalisation. 
%We did not include the 10-12 keV rest-frame band for the normalization because the background level at such energies in each spectrum is higher.

To minimise the effect of the Al spike at 1.5 keV in the spectra from XMM CDFS (discussed in Sect. \ref{parent_survey}), we ignored the bins corresponding to rest-frame energy in the range 1.4-1.6 keV observed frame (compensating for the lost flux by interpolating linearly to the full band) when estimating the continuum for the normalisation and also in the final average of both the real and the simulated data.
 
We binned the spectra using an energy grid with 100 eV-wide bins and we finally averaged them. The errors were estimated as the dispersion of the fluxes around the average.

We have found that if we use instead error bars obtained from error propagation of the individual source bins, as we did in \cite{falocco2012}, the combined error bars were smaller, resulting in very high reduced $\chi^2$ values. We believe that this is because, with these new data with more net counts, the intrinsic spectral dispersion dominates over the counting noise.  
%%   \begin{figure}\label{fig:z}
%%    \includegraphics[width=8cm,angle=0]{figures/norm_2_5_8_10_S/N10_S/N15.ps}
%% \caption{Distribution of the fluxes in 2-5 and 8-10 keV of the sample }
%%   \end{figure}

\subsection{Simulations of the continuum}\label{simu_cont}

We have estimated the expected shape of the underlying continuum in our spectra using comprehensive simulations.  

We simulated the spectrum of each source 110 times using the best-fit parameters of the simple
absorbed power law model from the individual fits to the real spectra
 (see Sect. 2.1). To each of these 110 simulated
samples we applied the same method as the one used for the observed
sample (spectral fitting, unfolding to correct for the instrument's energy-dependent response, correcting for
Galactic absorption, shift to rest-frame, normalising, averaging and calculation of errors from the dispersion). After this, we
adopted as the underlying continuum the median of the simulated
continua. The median is preferred over a simple average because it is a more robust estimate of the central value of a distribution.
%\citep{numericalrecipes}.

  \begin{figure}
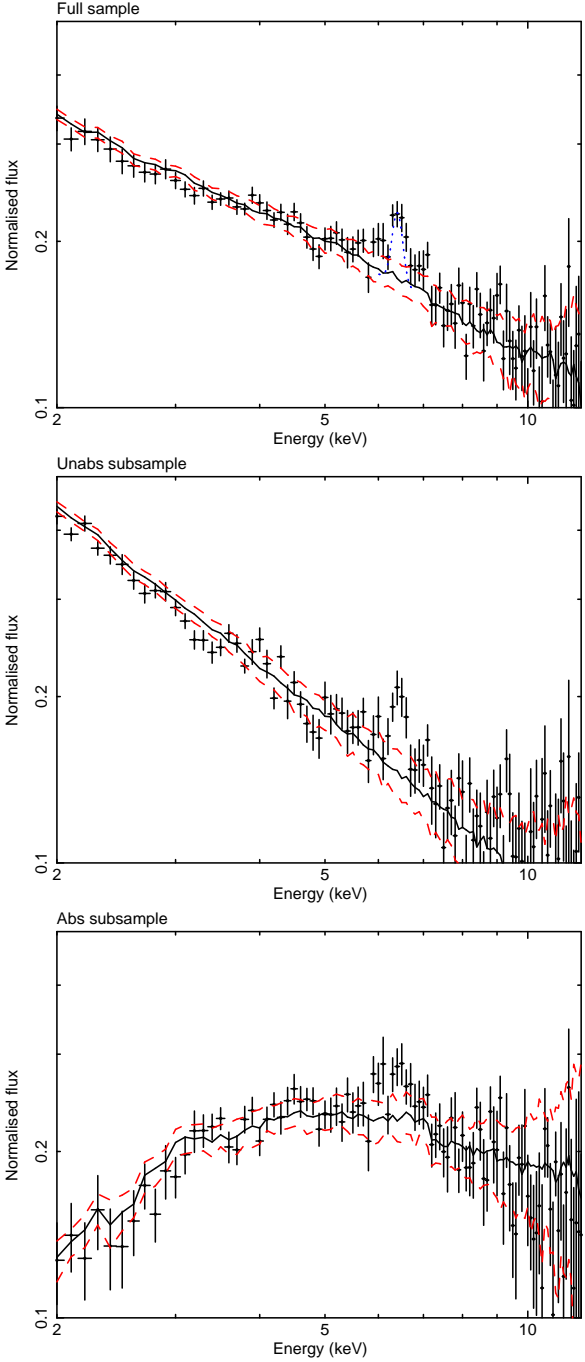

 \includegraphics[width=6cm,angle=-90]{figures/mediana_ave_no_each_clip_data_average_all_68_pl_grezzo_bin01.ps}\\
  \includegraphics[width=6cm,angle=-90]{figures/mediana_ave_no_each_clip_data_average_unabs_68_pl_usingerrors.ps}\\
  \includegraphics[width=6cm,angle=-90]{figures/mediana_ave_no_each_clip_data_average_abs_68_pl_usingerrors.ps}
\caption{Average observed spectrum (data points) with its simulated continuum (continuous line) and the $1\sigma$ confidence levels of our simulations (dashed lines). From top to bottom: full sample,{ Unabs subsample, Abs subsample}. The dotted line in the top panel represents the intrinsic resolution at 6.4 keV obtained from the simulations of the iron line described in Sect. \ref{simu_line}.}\label{fig:all_2_5_8_10}
  \end{figure}

%%   \begin{figure}
%%  \includegraphics[width=6cm,angle=-90]{figures/mediana_ave_no_each_clip_data_average_uncertain_68_pl.ps}\\
%% \caption{Average observed spectrum (data points) with its simulated continuum (continuous line) and the $1\sigma$ confidence levels of our simulations (dashed lines). Sample of sources with unconstrained absorption}\label{fig:all_2_5_8_10}
%%   \end{figure}

  \begin{figure}
   \includegraphics[width=6cm,angle=-90]{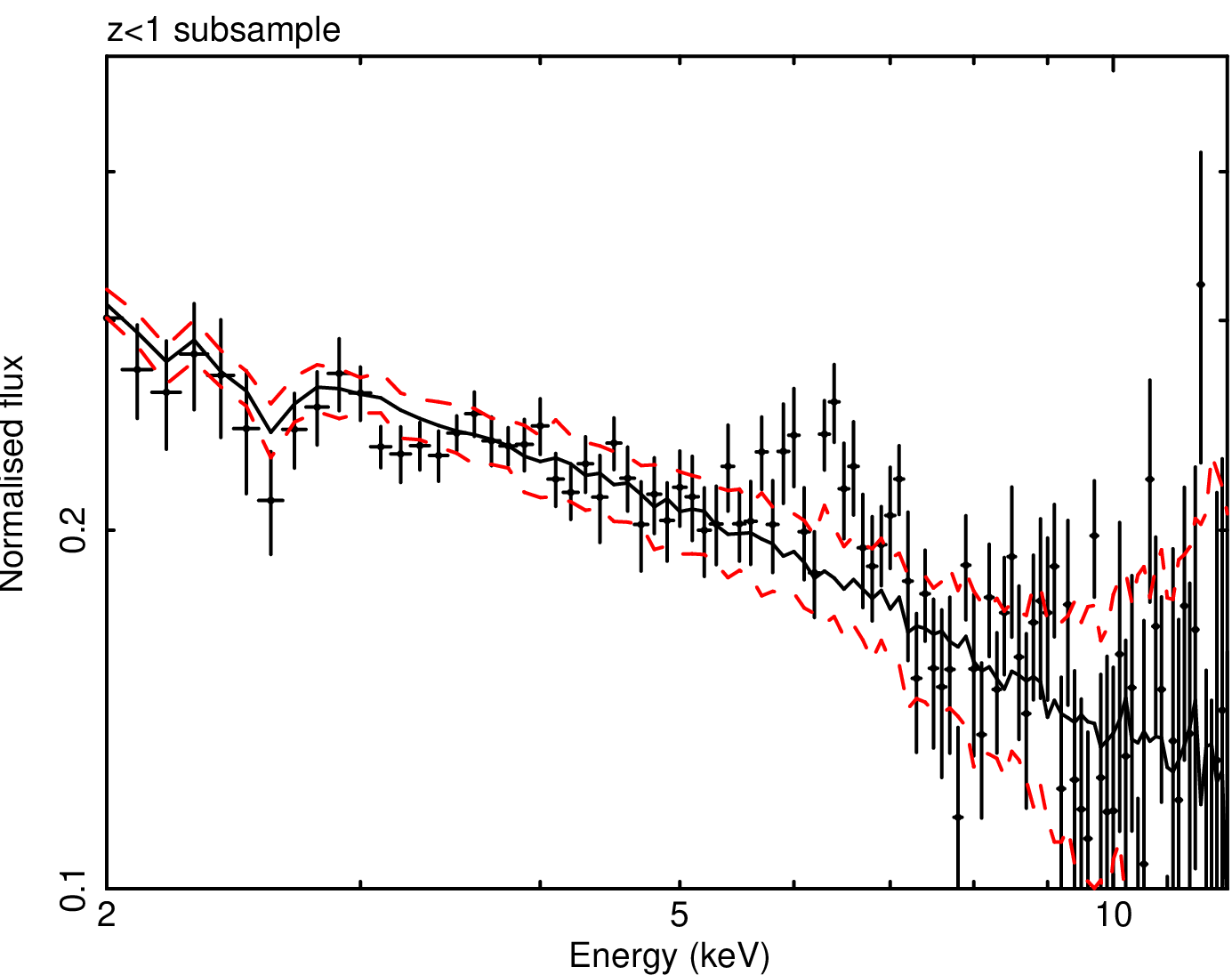}\\
  \includegraphics[width=6cm,angle=-90]{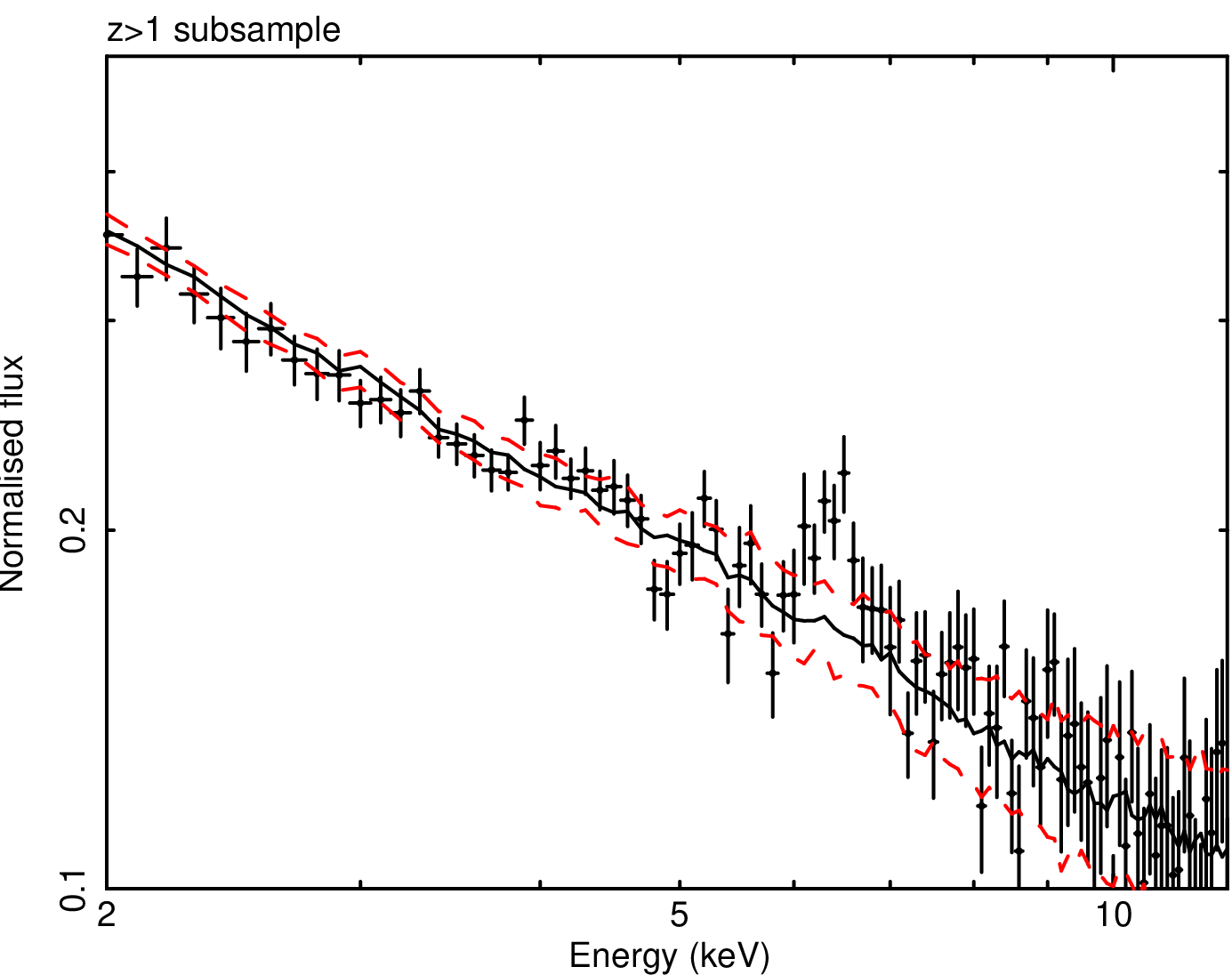}
\caption{Average observed spectrum with its simulated continuum and the $1\sigma$ confidence lines. Top panel: z$<$1  subsample, bottom panel: z$\geq$1 subsample. Symbols and line styles as in Fig. \ref{fig:all_2_5_8_10}.}\label{fig:z_2_5_8_10}
  \end{figure}

  \begin{figure}
   \includegraphics[width=6cm,angle=-90]{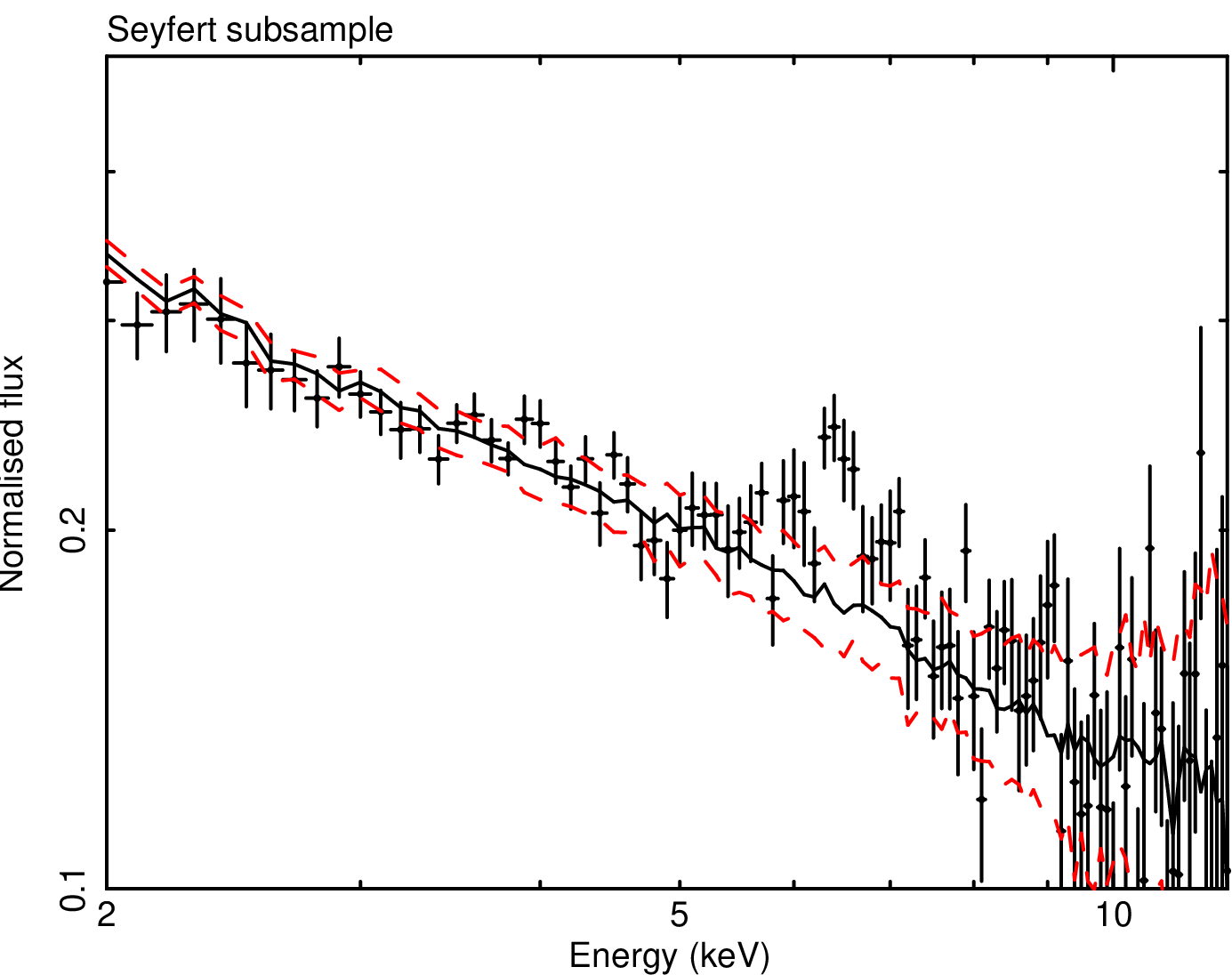}\\
  \includegraphics[width=6cm,angle=-90]{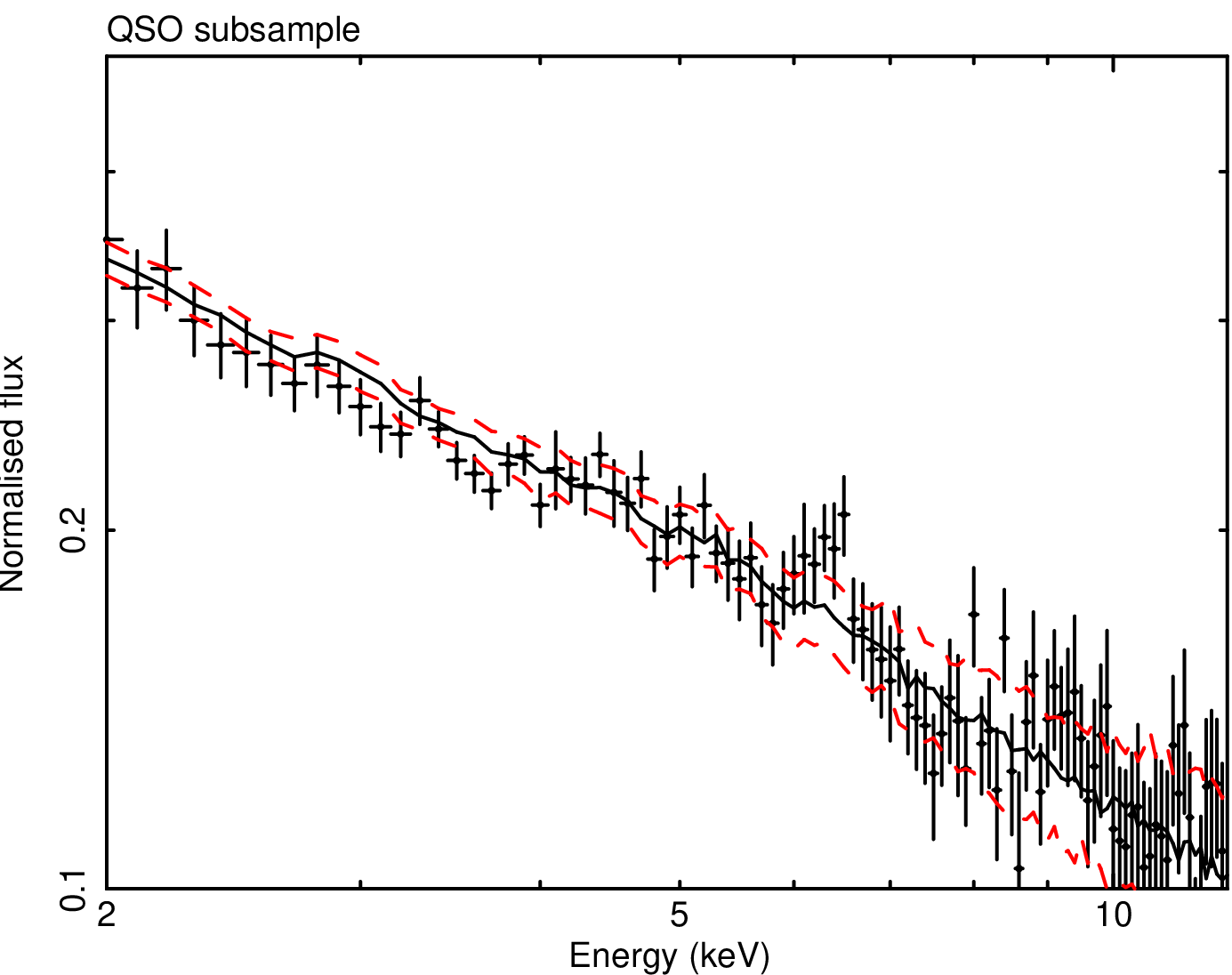}
\caption{Average observed spectrum with its simulated continuum and the $1\sigma$ confidence lines. Top panel: Sey subsample, bottom panel: QSO subsample. Symbols and line styles as in Fig. \ref{fig:all_2_5_8_10}.}\label{fig:lum_2_5_8_10}
  \end{figure}

We present in Figs. \ref{fig:all_2_5_8_10}, \ref{fig:z_2_5_8_10}, \ref{fig:lum_2_5_8_10} our average spectra together with their corresponding simulated continua. In those figures we present: the average observed spectrum (data points), our simulated continuum (continuous line) and the $1\sigma$
confidence lines of our continuum simulations (dashed lines). The latter were estimated from the 68\% percentiles around the median. 

An intense iron line characterises the observed spectra of
the sample and all the
subsamples that we constructed. Its profile appears visually broader than the instrumental resolution, in the full sample (shown in the top panel of Fig. \ref{fig:all_2_5_8_10}).

\subsection{Simulations of the Fe Line}\label{simu_line}
The spectral resolution of the X-ray detectors and the averaging process can widen any narrow spectral features in the X-ray spectra making them to appear artificially broader than they actually are. The resulting effective spectral dispersion changes with energy and, consequently, our ability to resolve spectral features will change along the spectrum. 
To properly study the iron line profile in the XMM CDFS, we need to estimate our effective spectral dispersion in the energy band we are analysing.
To do this, we ran high S/N simulations of unresolved ($\sigma=0$) Gaussian emission lines centred at rest-frame energies between 1 and 10 keV, in steps of 1 keV with input EW of 200 eV (one simulation for each source -at its observed redshift-, and for each energy). The input continuum model was a powerlaw with $\Gamma=1.9$ without any intrinsic absorption. We corrected the spectra for detector response (modelling them with a powerlaw with photon index 1.9) as done in \cite{corral2008} and \cite{falocco2012}; then we applied the same treatment described in Sect. \ref{method}. 
We then studied how the width of the simulated lines changes with the central energy. In addition, in analogy with what we did in \cite{falocco2012}, we simulated lines centred at 6.4 keV and 6.9 keV to estimate the resolution at the peak energy of the line from neutral and ionised iron.
We recovered the input parameters of the continuum very well, reproducing a powerlaw continuum with $\Gamma \sim$1.9, while the Gaussian lines appear broadened by $\Sigma\sim$ 100 eV. 
A similar value of the spectral resolution given by the stacking procedure and the X-ray instruments was found in the study of the \textsl{XMM-Newton} X-ray spectra in the COSMOS sample \citep{iwasawa2011}.

%with $\sigma\sim 113~eV$ at 6.4 keV and $\sigma\sim 117~eV$ at 6.9 keV.\\

We studied the width of the average simulated lines as a function of their centroid energies: the result shows, for the full sample, $\Sigma =119~ \rm eV ~\left(\frac{E}{6~ \rm keV}\right)^{0.34}$. The values of the $\Sigma$ obtained in this work include the contributions from the spectral dispersion of the instruments, {$\Sigma \sim$ 55 eV} \citep{haberl2002}, as well as the widening introduced by the averaging method. The parameters of this $\Sigma(E)$ dependence (the slope and the $\Sigma$ at 6 keV) for the full sample and its subsamples are in Table \ref{tpropertiesS/N15}. This energy-dependent instrumental response has been taken into account in the spectral fitting of the average spectra.

We emphasise that this limited spectral resolution essentially means that any features narrower than about 0.1 keV are very likely due to statistical fluctuations.

\section{Results}\label{results}
We discuss in this section the results of our analysis of the average \textsl{XMM-Newton} X-ray spectra of several subsamples in the XMM CDFS, estimating in particular the iron line significance and shape. A new way to estimate the line EW independent of the fits to the average spectra has been applied and it is described in \ref{model_ind}. 
We describe the detailed spectral fits to the data in the Sect. \ref{specfits}.

\subsection{Model-independent estimate of the line EW}\label{model_ind}

%% \begin{table}
%% \caption{\label{t_model_independent} Number of simulations over the total with a lower flux than the flux of the average observed spectrum (see text). The bandwidths are expressed in keV.}
%% \begin{tabular}{llllll}
%% Energy band & {\it narrow}  &{\it neutral} & {\it bulk}  & {\it ionized} \\
%% Sample & 6.2-6.6  &6.0-6.8 & 5.5-7.  & 6.8-7.0 \\
%% \hline
%% Full       & 110/110  & 110/110& 110/110     & 110/110  \\
%% \hline
%% $\log{\left( N_{\rm H}\right)}<21.5$     & 110/110 & 110/110& 110/110     &110/110  \\
%% $\log{\left( N_{\rm H}\right)}\geq21.5$       & 110/110 & 110/110 & 110/110  & 101/110 \\
%% \hline
%% z$<$1      & 110/110& 110/110& 110/110     & 110/110  \\
%% z$>$1  & 110/110& 110/110& 110/110   & 109/110\\
%% \hline
%% $\log{\left( L\right)}<$ 44    & 110/110 & 110/110& 110/110     & 110/110 \\
%% $\log{\left( L\right)}\geq$ 44  &110/110  & 110/110& 107/110     &  30/110 \\
%% \end{tabular}
%% \end{table}

We compared the features found in our average observed spectrum with the average simulated spectra to estimate the line EW (see Table \ref{teqw}). 
%To make an estimation of how the observed feature deviate from the expected (simulated) continua, in analogy of what we did in \cite{falocco2012}, we found out the fraction of simulations with a flux in a given range lower than that of the observed spectrum. 
To do this, we used the following energy ranges: 5.5-7 keV (that we call, for convention, {\it bulk}), 6-6.8 keV (called {\it neutral}), 6.2-6.6 keV (called {\it narrow}), 6.8-7 keV (called {\it ionised}), chosen to represent several spectral features from relativistically-broadened neutral Fe-K lines (5.5-7. keV), and from narrow neutral Fe lines (6.2-6.6 keV), to symmetrically broadened lines (6.0-6.8 keV) to lines from H-like Fe (6.8-7.0 keV), taking our spectral resolution ($\Sigma\sim0.1$~keV) into account. We did not considered an interval centred at 6.7 keV to represent the line from He-like Fe because the average spectra do not visually display this feature (see Figs. 6-8).

%%  An excess in the observed data between 6.2 and 6.6 keV (the range where the putative narrow line is expected) is found with respect to all our simulations, for the full sample and its subsamples (see Column 2 of Table 2 for details). 

%% If the broadest range is considered, i.e. from 5.5 to 7.0 keV (the range where the relativistic line component is expected to extend) the excess is found with respect to 100\% of the simulations in the majority of the subsamples that we constructed, except for the $\log L \geq 44 $ subsample where the fraction of simulations is 97\% (see Col. 3 of Table 2 for details). 
%% An excess in our observed data is found also in the range 6.8-7.0 keV, in the majority of the cases with the exception of the Abs subsample (where it is found in the comparison with 101 of our simulations) and the $\log L \geq 44$ subsample (where the fraction of simulations is only 30/110). 

We estimated the EW of the iron line using the 110 simulated continua and used the same energy bands described above.
For each simulation we calculated the integral:
%\begin{equation}
%EW=$\sum_{k=1}^N k^2$
%EW=$ \int_{E_1}^{E_2} dE \frac{T(E)-C(E)}{C(E)}\sim 2 \sum_{E_{1}}^{E_{2}} \Delta (E_i)\frac{T(E_{i})-C(E_{i})}{C(E_{i})}$
EW=$ \int_{E_1}^{E_2} dE \frac{T(E)-C(E)}{C(E)}\sim  \sum \Delta (E_i)\frac{T(E_{i})-C(E_{i})}{C(E_{i})}$
%\end{equation}
where $\Delta(E_i)$ is the width of the bin with centre $E_i$; $T(E_i)$ represents the average observed spectrum, $C(E_i)$ the average continuum.
The values of the median of the EW for each subsample and energy range are shown in Table \ref{teqw}. The confidence levels in that Table are calculated from the 68\% percentiles around the corresponding medians. 

There is evidence for a significant decrease of the EW of the narrow line with increasing average luminosity, in agreement with the 'Iwasawa-Taniguchi effect' \citep{iwasawa94}. No significant trend with redshift has been found.
 The EW estimated in the 'bulk' range is higher for the Unabs subsample than for the Abs subsample, as expected (because unabsorbed AGN are seen preferentially face-on), we should note that the N$\rm _{H}$ values covered by the absorbed sample that we defined in this work are too low to affect significantly the continuum above 6 keV, thus the iron line region. The same consideration should be applied to understand also the lower EW estimated in the 'neutral range' for the Abs subsample, despite a higher EW of the narrow core expected when absorption is significant. 
%We should note also that the EW estimated in the 'neutral' range for the Absorbed sample is lower than the EW estimate of the Unabsorbed one.
\begin{table*}
  \caption{\label{teqw}Median of the EW calculated in different energy ranges using the simulations. }
 \centering
\scalebox{1.1}{
 \begin{tabular}{llllll}

%% Sample & 5.5-7. & 6.-6.8  & 6.2-6.6  &  6.3-6.5  & 6.8-7. \\

Energy band & {\it Bulk} & {\it Neutral}  & {\it Narrow}  & {\it Ionised} \\

sample & 5.5-7. & 6.0-6.8  & 6.2-6.6  &  6.8-7. \\
\hline 
%% Full & $203 \pm 40 $%_{-30} ^{+33} $ 
%% & $160  \pm 31 $%_{-23} ^{+21}$ 
%% &    $117 \pm 24 $%_{-16}^{+14} $ 
%% %&  $88 \pm 19 $%_{-13} ^{+15} $ 
%% & $26 \pm 19  $%_{-11} ^{+15}$ 

Full & $203_{-30} ^{+33} $ 
& $160_{-23} ^{+21}$ 
&    $117_{-16}^{+14} $ 
%&  $88_{-13} ^{+15} $ 
& $26_{-11} ^{+15}$ 
\\
\hline
%$\log{\left( N_{\rm H}\right)}< 21.5$
%% Unabs & $230 \pm 54 $%_{-33} ^{+36} $ 
%% & $184  \pm 41 $%_{-35} ^{+30}$ 
%% &    $145 \pm 32 $%_{-22}^{+27} $ 
%% %&  $107 \pm 26 $%_{-20} ^{+22} $ 
%% & $34 \pm 25  $%_{-17} ^{+22}$ 
{Unabs} & $252_{-38}^{+48} $ 
& $194_{-38}^{+42}$ 
&    $156_{-26}^{+40} $ 
%&  $107 \pm 26_{-20} ^{+22} $ 
& $38_{-24} ^{+18}$ 
\\
\hline

%$\log{\left( N_{\rm H}\right)}\geq 21.5$ 
%% Abs  & $178 \pm 56 $%_{-50} ^{+48} $ 
%% & $132  \pm 44 $%_{-30} ^{+31}$ 
%% &    $76 \pm 30 $%_{-23}^{+29} $ 
%% %&  $64 \pm 27 $%_{-20} ^{+19} $ 
%% & $18 \pm 27  $%_{-18} ^{+13}$ 

{Abs}  & $192_{-26} ^{+26} $ 
& $150_{-30} ^{+38}$ 
&    $86_{-46}^{+46} $ 
%&  $64_{-20} ^{+19} $ 
& $18^{+16}_{-12}$ 
\\
\hline
%% z$<$1  & $254 \pm 66 $%_{-53} ^{+50} $ 
%% & $166  \pm 49 $%_{-34} ^{+38}$ 
%% &    $119 \pm 39 $%_{-29}^{+28} $ 
%% %&  $92 \pm 29 $%_{-23} ^{+23} $ 
%% & $37 \pm 29  $%_{-18} ^{+19}$ 
%% \\
z$<$1  & $254_{-53} ^{+50} $ 
& $166_{-34} ^{+38}$ 
&    $119_{-29}^{+28} $ 
%&  $92 \pm 29 $%_{-23} ^{+23} $ 
& $37_{-18} ^{+19}$ 
\\
\hline

%% z$\geq $1 & $176 \pm 54 $%_{-47} ^{+37} $ 
%% & $160  \pm 41 $%_{-31} ^{+28}$ 
%% &    $118 \pm 31 $%_{-22}^{+20} $ 
%% %&  $88 \pm 25 $%_{-19} ^{+17} $ 
%% & $21 \pm 27  $%_{-17} ^{+16}$ 
%% \\

z$\geq $1 & $176_{-47} ^{+37} $ 
& $160_{-31} ^{+28}$ 
&    $118_{-22}^{+20} $ 
%&  $88 \pm 25 $%_{-19} ^{+17} $ 
& $21_{-17} ^{+16}$ 
\\
\hline
%$\log{\left( L\right)}<44$ 
%% Sey  & $283 \pm 59 $%_{-48} ^{+51} $ 
%% & $213  \pm 47 $%_{-37} ^{+26}$ 
%% &    $148 \pm 34 $%_{-21}^{+22} $ 
%% %&  $112 \pm 29 $%_{-22} ^{+17} $ 
%% & $47 \pm 27  $%_{-20} ^{+20}$ 
%% \\
  Sey  & $283_{-48} ^{+51} $ 
& $213_{-37} ^{+26}$ 
&    $148_{-21}^{+22} $ 
%&  $112 \pm 29 $%_{-22} ^{+17} $ 
& $47_{-20} ^{+20}$ 
\\
\hline

%$\log{\left( L\right)}\geq44$ 
%% QSO & $71 \pm 53 $%_{-29} ^{+40} $ 
%% & $87  \pm 40 $%_{-25} ^{+25}$ 
%% &    $66 \pm 32 $%_{-15}^{+21} $ 
%% %&  $54 \pm 23 $%_{-15} ^{+16} $ 
%% & $<15  $%_{-15} ^{+16}$ 
%% \\
 QSO & $71_{-29} ^{+40} $ 
& $87_{-25} ^{+25}$ 
&    $66_{-15}^{+21} $ 
%&  $54 \pm 23 $%_{-15} ^{+16} $ 
& $<32$ 
\\

\end{tabular}
}
 \tablefoot{The errors include both statistical errors propagated through the stacking process as well as the uncertainties in the continuum from the 110 simulations. The range is expressed in keV and the EW in eV.}
\end{table*}

\subsection{Spectral fits to the full sample and subsamples}
\label{specfits}
To characterise the spectral features of our average spectra and to understand their physical origin, we fitted them in the full rest-frame 2-12 keV band (ignoring the E$<$2 keV rest-frame band in this analysis because we are not interested here in studying the soft excess).
%We excluded the spectral band with E $>$ 12 keV because the background level grows with the energy and noise can be dominant at high energies. 

In our analysis of the deep \textsl{Chandra} fields of \cite{falocco2012} the average simulated spectrum was employed as a continuum in the spectral fits. 
That approach does not take the uncertainty in the shape of the underlying continuum into account.
In the current work, we preferred to construct an empirical continuum with a flexible, smooth shape.
 To take the instrumental line broadening discussed in Sect. \ref{simu_line} into account, we smoothed the input spectral model with a Gaussian convolution model (\texttt{gsmooth} in \texttt{xspec}). This model smooths the spectral features with a Gaussian with a variable sigma, which changes as a function of the energy (the powerlaw slope and the $\Sigma$ found in our simulations of the unresolved lines in Sect. \ref{simu_line} and summarised in Table \ref{tpropertiesS/N15} are used), hence, all the parameter values given below are 'intrinsic' values, i.e., with the instrumental/method resolution already subtracted.

Our continuum is an absorbed powerlaw plus a Compton reflection component (both represented by a \texttt{pexrav} model in \texttt{xspec}), fitted individually to the average spectrum of the full sample and each subsample, excluding initially the 5-7.2 range (to avoid the iron feature region).

We reintroduced the channels between 5 and 7.2 keV and we show in Col. 2 of Table \ref{tresults} the corresponding $\chi^2$.
The simple continuum model fits the overall continuum reasonably well in all cases, as the values of $\chi^{2}_{\nu}$ indicate. We found an excess in the iron line region that we model with a simple phenomenological Gaussian (see \ref{gaussianfits}) and with a more realistic model (see \ref{broadLines}). The significance of any fit improvement is calculated, from the $\chi^2$, using $P(\Delta\nu,\Delta\chi^2)={\mathbf P}(\Delta\nu/2,\Delta\chi^2/2)$, where ${\mathbf P}(\Delta\nu/2,\Delta\chi^2/2)$ is the incomplete gamma function \citep{numericalrecipes}. 

\subsubsection{Iron line fitting with phenomenological models}
\label{gaussianfits}
After having re-introduced the bins between 5 and 7.2 keV, that contain the iron line, we wanted to model the excess found in the iron line region. The first model we used is a simple phenomenological Gaussian.

We set the parameters of the Gaussian as follows:
\begin{itemize}
\item Fixed $ \sigma$=0 and fixed centroid energy at 6.4 keV: this
  allows us to estimate the significance of the narrow component of a neutral
  Fe K line (hereafter '$E_{fixed}-\sigma_{fixed}$', in the first line of each sample in Table \ref{tresults})
\item Fixed centroid energy at 6.4 keV and free $\sigma$: it studies 
  the significance of a possible broad component and
estimates its width (hereafter '$E_{fixed}-\sigma_{free}$', in the second line of each sample in Table \ref{tresults})
\item Free centroid energy and fixed $\sigma$=0: it considers the
  presence of a narrow Fe component centred at energies $\geq$6.4 keV and estimates its centroid
  energy (hereafter '$E_{free}-\sigma_{fixed}$', in the third line of each sample in Table \ref{tresults})
\item Free centroid energy and free $\sigma$: considers both
  iron emission centred at energies $\geq$ 6.4 keV and broad line emission (hereafter '$E_{free}-\sigma_{free}$', in the fourth line of each sample in Table \ref{tresults})
\end{itemize}

In Col. 4 of Table \ref{tresults} the probability of any fit improvement with $\sigma_{fixed}-E_{fixed}$ with respect to the fit with the continuum only is presented (in the first line of each sample). The probability of the improvement of $E_{free}-\sigma_{fixed}$, $E_{free}-\sigma_{fixed}$ and $E_{free}-\sigma_{free}$ with respect to $\sigma_{fixed}-E_{fixed}$ model is shown instead in the second, third, fourth line corresponding to each sample.

The $\sigma$ given in Table \ref{tresults} is intrinsic, since our effective spectral resolution has already been taken into account through the convolution model \texttt{gsmooth}.

{The values of the photon index of the average spectra are rather flat in all cases with the exception of the Unabs subsample. Fitting the simulated continuum of the full sample with the model \texttt{gsmooth(pha * pex)} we found continuum parameters consistent with those reported in Table }\ref{tresults} {($\Gamma=1.66\pm0.01$ for the simulated continuum). }

The values of the column density $\rm N_H$ obtained in the fits to the average observed spectra are moderate (see Table \ref{tresults} and Fig. 4 for details), recovering the moderate absorption in individual spectra (see Table \ref{tpropertiesS/N15}).
The significance of the emission line at 6.4 keV is $6.8\sigma$ in the full sample and always higher than 3.9 $\sigma$ in its subsamples (Col. 4 of Table \ref{tresults}). Leaving {the Gaussian width and/or the central energy} free in the spectral fitting is not required by our data in our full sample and the majority of the subsamples that we constructed. Given that the fit with free centroid energies gives output central energies equal to 6.4 keV, we do not add in the following analysis a further line from ionised iron. 

 {The QSO subsample has a worse statistical description with respect to the other subsamples (see the $\chi^2/dof$ in Table \ref{tresults}), due to relatively strong residuals above  about 8-9 keV rest-frame which cannot be accounted for by the actual modelling. Moreover, the  QSO subsample is characterised by a relatively low spectral dispersion and thus flux errors significantly smaller than the other subsamples. }

\subsubsection{Fits with physical models}\label{broadLines}

Having established in the previous Sections the presence of Fe K emission features in the average spectra of our full sample and subsamples, we will now use physically motivated models to characterise its shape. The iron line model is overlaid to the same continuum model considered in the previous section; its profile is modelled by a Gaussian with fixed energy at 6.4 keV and no intrinsic width described in the previous section (that represents fluorescence from material far from the central BH) and a \texttt{laor} component (representing fluorescence from the accretion disk {around a rotating BH}). The \texttt{laor} model used here was developed by \cite{laor} to test the accretion disk hypothesis and it assumes a {Kerr} metric, valid in an accretion disk around a {rotating BH}. Several other models are available in the literature but we chose to fit our average iron lines with {one of the simplest ones} because, as shown below, our data clearly does not warrant a more sophisticated modelling.
 We set the \texttt{laor} parameters making simple considerations about the geometry and the physics expected in our sources.

The model adopted assumes emission from an accretion disk with emissivity index {-3}, in the hypothesis that the continuum source is located in the disk axis and at low height from the disk surface.
Since our sample is dominated by unabsorbed sources that constitute {59} \% of the full sample (see Table \ref{tpropertiesS/N15}), we expect that the accretion disks are seen preferentially face-on, under the hypothesis that the accretion disk and the torus axes are aligned. 
The accretion disk is assumed to extend from {1.23} to {400} R$_g$ as it is the case for a disk around a {maximally rotating BH}.
Hereafter, we will refer to the continuum plus the fluorescent lines (both narrow Gaussian and \texttt{laor}) as to the 'two-component model'. 

\begin{figure}
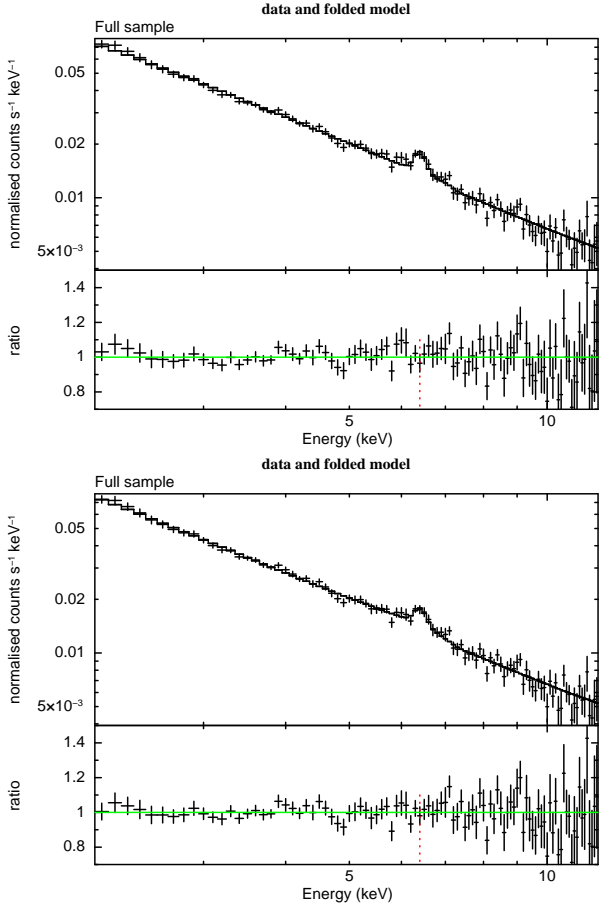

\includegraphics[width=6cm,angle=-90]{figures/all_errdev_gsmo_pha_pex_fix_fix.ps}\\ 
\includegraphics[width=6cm,angle=-90]{figures/all_errdev_gsmo_pha_pex_fix_fix_laor_fix.ps}
\caption{Full sample. Top panel: Fit with $E_{fixed}-\sigma_{fixed}$. Bottom panel: fit with the 'two-component model'. \label{fits_full}} 
  \end{figure}

%% \begin{figure}
%% %\includegraphics[width=6cm,angle=-90]{figures/abs_errdev_gsmo_pha_pex_fix_fix.ps}\\ 
%% %\includegraphics[width=6cm,angle=-90]{figures/abs_errdev_gsmo_pha_pex_fix_fix_disk_fix.ps}\\
%% \includegraphics[width=6cm,angle=-90]{figures/abs_errdev_gsmo_pha_pex_fix_fix_disk_fix.ps}\\
%% \caption{Abs sample. Fit with the 'two-component model'\label{fits_abs}} 
%%   \end{figure}

\begin{figure}
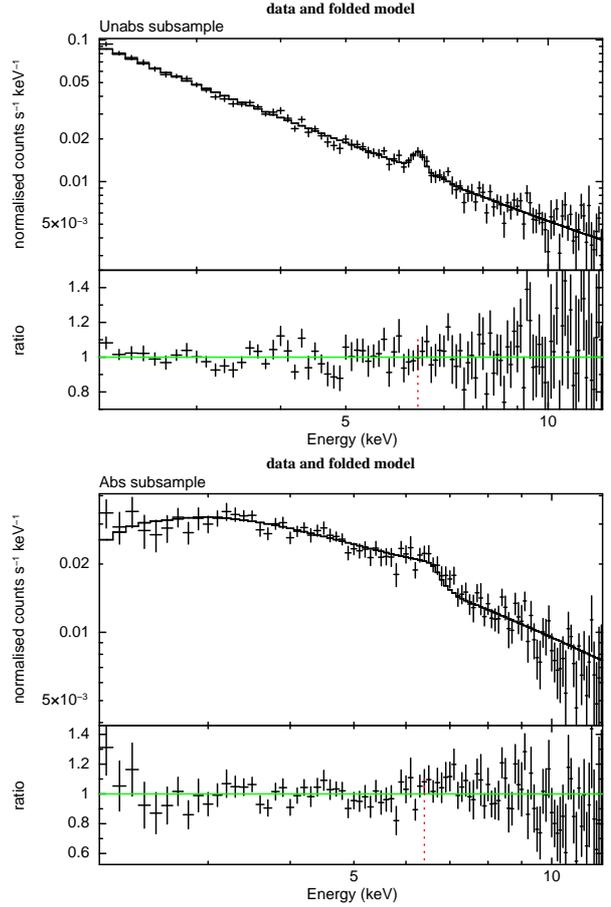

\includegraphics[width=6cm,angle=-90]{figures/unabs_errdev_gsmo_pha_pex_fix_fix_laor_fix.ps}\\
\includegraphics[width=6cm,angle=-90]{figures/abs_errdev_gsmo_pha_pex_fix_fix_laor_fix.ps}
\caption{Fit with the 'two-component model'.{ Top panel: fit of the Unabs subsample. Bottom panel: fit to the Abs subsample}.%Top panel: Fit with $E_{fixed}-\sigma_{fixed}$. Bottom panel: fit with the 'two-component model'\label{fits_unabs}
\label{fits_unabs}} 
  \end{figure}

%\begin{figure}
%\includegraphics[width=6cm,angle=-90]{figures/Llo44_errdev_gsmo_pha_pex_fix_fix.ps}\\ 
%\includegraphics[width=6cm,angle=-90]{figures/Llo44_errdev_gsmo_pha_pex_fix_fix_disk_fix.ps}\\

%\caption{Sey subsample. Top panel: Fit with $E_{fixed}-\sigma_{fixed}$. Bottom panel: fit with the 'two-component model'\label{fits_Llo44}} 
%\caption{Sey subsample. Fit with the 'two-component model' \label{fits_Llo44}} 
 % \end{figure}

\begin{figure}
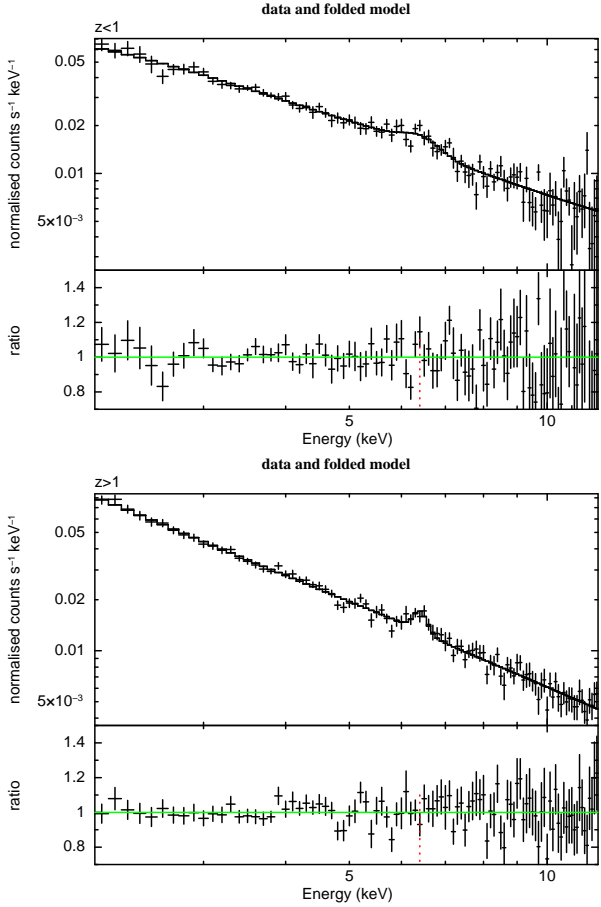

\includegraphics[width=6cm,angle=-90]{figures/zmin1_errdev_gsmo_pha_pex_fix_fix_laor_fix.ps}\\
\includegraphics[width=6cm,angle=-90]{figures/zma1_errdev_gsmo_pha_pex_fix_fix_laor_fix.ps}
\caption{Fit with the 'two-component model'. { Top panel: subsample with z$<$1. Bottom panel: subsample with z$\geq$1.} \label{fits_zma1}} 
  \end{figure}

\begin{figure}
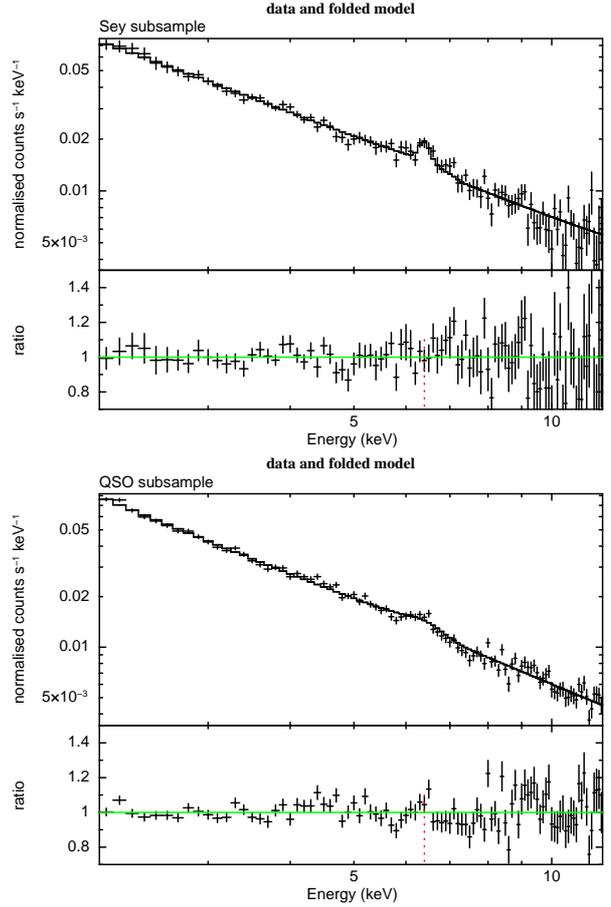

\includegraphics[width=6cm,angle=-90]{figures/Llo44_errdev_gsmo_pha_pex_fix_fix_laor_fix.ps}\\
\includegraphics[width=6cm,angle=-90]{figures/Lhi44_errdev_gsmo_pha_pex_fix_fix_laor_fix.ps}
\caption{Fit with the 'two-component model'. {Top panel: Sey subsample. Bottom panel: QSO subsample.} \label{fits_Lhi44}\label{fits_Llo44}}
%\caption{QSO subsample. Top panel: Fit with $E_{fixed}-\sigma_{fixed}$. Bottom panel: Fit with a narrow gaussian at 6.4 keV plus \texttt{diskline}\label{fits_Lhi44}} 
  \end{figure}

%% \begin{figure}
%% %\includegraphics[width=6cm,angle=-90]{figures/zmin1_errdev_gsmo_pha_pex_fix_fix.ps}\\ 
%% %\includegraphics[width=6cm,angle=-90]{figures/zmin1_errdev_gsmo_pha_pex_fix_fix_disk_fix.ps}\\

%% \caption{Subsample with z$<$1. Fit with the 'two-component model' \label{fits_zmin1}} 
%%   \end{figure}

The fits to our data and the ratios between data and model are shown in Figs. from \ref{fits_full} to \ref{fits_zma1}; the corresponding fit results are given in Table \ref{tfix_fix_laor}. {In the figures, a dotted vertical line has been placed at 6.4 keV energy for reference.} In Fig. \ref{fits_full}, we present, for comparison, the fits to the narrow Gaussian centred at 6.4 keV (the $E_{fixed}-\sigma_{fixed}$ described in the previous paragraph) in the top panel and the fits to the two-component model in the second panel. The fit is visually very similar in both cases. This impression is confirmed by looking at the first row of Table \ref{tfix_fix_laor} that shows that, although the fit is better including both components, the improvement is statistically significant only at $2.1 \sigma$. 

The formal 90\% confidence levels on the EW of both components almost reach zero in all cases, this is a consequence of the very strong coupling between their fluxes, as will be discussed below (see also Fig. \ref{contour_abs}).\\
We show in the last two columns of Table \ref{tfix_fix_laor} the $\chi^2/\nu$ and the EW of the relativistic component obtained after setting the flux of the narrow component to zero.

\subsubsection{Fits with comprehensive models}\label{comprehensive}

{We finally made a test using a full-inclusive model that considers the entire set of spectral components expected in AGNs: the iron K$_{\alpha}$ at 6.4 keV, the Fe K}$_{\beta}${ at 7.05 keV, the Ni K}$_{\alpha}${ at 7.47 keV, the Compton shoulder to the Fe K}$_{\alpha}${ (approximated to be a Gaussian at 6.3 keV with width 35 eV) and the Compton reflection component of the continuum. This last component describes reflection in a slab geometry.
The full-inclusive model, namely \texttt{pexmon}, was presented by \cite{nandra2007}. We used two \texttt{pexmon} components to consider reprocessing from material close to the central engine and far away from it. If the emission comes from the innermost regions, the \texttt{pexmon} will be smeared by the relativistic effects due to the proximity of the central SMBH. To account for these effects we convolved one \texttt{pexmon} component with the relativistic kernel \texttt{kdblur2} which smears both the fluorescent lines and the Compton reflection. The resulting model will be absorbed by circumnuclear material (\texttt{pha} model). The second \texttt{pexmon} component, instead, will not be smeared or absorbed because it originates in external regions.}

{
The convolution model \texttt{kdblur2} used here has a similar set-up
to that of \cite{nandra2007} to allow a useful comparison with iron lines
detected in individual spectra of AGN with high spectral quality. The
model considers two emissivity laws, with slopes of 0 and 3, and a
break radius which defines the border between the two regions
initially fixed to 20 } $R_g$ { for simplicity. We considered an accretion
disk around a maximally rotating BH, extended from 1.235 to 400} $R_g${, to allow a comparison with the results presented in the previous
paragraph. We then thawed the inner radius and the
break radius, but the results do not change significantly: in the first case we found an inner radius of $<62R_g$}
%$11.00^{51.00}_{9.77}R_g$ 
with $\chi^2/dof=83.9/93$
{; in the second
case we obtained a break radius of $<131R_g$} 
%$25.00_{-23.77}^{+106.00}R_g$ 
with $\chi^2/dof=84.1/93$.  

{The comprehensive model just described allows a description of the continuum more accurate than the phenomenological and the 'two component' model. The values of the photon index increase from 1.63$^{+0.03}_{-0.09}$ reported in Table \ref{tresults} to the 1.71$\pm$0.06 value of Table \ref{tpexm}, although they remain compatible within their error bars. The reflection factors and their 90\% confidence limits, -0.35$\pm$0.18 and 0.45$^{+0.38}_{-0.35}$, are both different from zero, suggesting that the two components have been found (see Sect. \ref{discussion_comprehensive}).}

\section{Discussion}\label{discussion}
We discuss in this Section the results obtained with the simulations, with the fits to a phenomenological model, and with the fits to physically-motivated models.
\subsection{Discussion based on the simulations}\label{discussion_simu}
Accurate modelling of the continuum is critical in determining the iron line properties of AGN and for this reason we developed a new method to do it independently of the spectral fitting.
The EW discussed here are estimated with respect to a continuum integrated between 5.5 and 7. keV in the full sample (see Sect. \ref{model_ind}).
We expect at least two Fe line components within 5.5 and 7.0 keV: a narrow Fe line from neutral material with EW of 50-100 eV ubiquitously seen both in nearby and distant AGN \citep{nandra2007,corral2008,corral2011} and a broad line component with a distribution spanning from virtually zero to 300 eV and a median around 170 eV, as in the Seyferts sample of \cite{delacalle2010}. 
The EW of 203$^{+33}_{-30}$ eV {found in the full sample is therefore entirely consistent with having both components and, most importantly, not consistent with having only a narrow component (which is invariably smaller than 100 eV in AGN at low redshift).}
The estimates of the total EW in the subsamples also are compatible with the expected values, because the EW range from 71 eV (in the QSO subsample) to 283 eV (in the z$<$1 subsample). 
%{As far as the z$<1$ subsample is concerned, we note that these values are not only consistent with having a double component (a narrow and a relativistic iron line) but, most importantly, not consistent with having only a narrow component (which is invariably smaller than 100 eV at low redshift). }

In addition, the EW estimated with our simulations are generally higher when using a broader spectral window, which is compatible with the presence of a broad emission line.

\subsection{Discussion based on phenomenological models}\label{discussion_phenomenol}
We considered a simple phenomenological Gaussian to fit the
  iron line and calculated the EW considering a
  monochromatic continuum at 6.4 keV. In these fits, we found that it
  is required by the data with high significance: from the spectral fitting to a simple
Gaussian, the improvement of adding the Gaussian to the continuum is
at $1-P(\Delta\nu,\Delta\chi^2)=9\times10^{-12}$ in the full sample
(see Table \ref{tresults}), value that corresponds to a $6.8\sigma$
confidence level.  The EW for the narrow line, is 95$\pm$22 eV for the
full sample and lays between {77}$\pm$35 eV (for the Abs subsample) and 167$\pm$67 eV (for the z$<1$ subsample): again, such values are consistent with those of
fluorescent lines from material far away from the central BH.

 Allowing for a broad neutral line ($E_{fixed}$-$\sigma_{free}$ second row of each subsample in Table \ref{tresults}), the errors on the line physical width could not be constrained in the full sample and in the majority of its subsamples: the upper limit of $\sim300$ eV found in the full sample indicates that we cannot exclude the presence a broad line component in our detected lines although we are not able to constrain its properties.
%{, basically due to a strong degeneracy between the Compton Hump and the iron line component. For this reson, considering that we obtain, for $R$, an upper limit consistent with one in the majority of the cases studied here, we repeated the fit of the full sample with $E_{fixed}$-$\sigma_{free}$ fixing it and we found a $\sigma=91^{+220}_{-91}$ and an EW=$107 \pm 60$ eV, consistent with the value reported in Col. 9 of table (with similar continuum parameters)} \ref{tresults}. 
Looking at Table \ref{tresults}, Col. 4, the subsamples where a significance higher than $3\sigma$ is found in the fit with $E_{fixed}-\sigma_{free}$ (second row of each subsample) are the Abs subsample and QSO subsample. In both cases, there is a degeneracy between the iron line and the reflection component.
%this is achieved at the expense of reducing the relative reflection parameter to R$\sim$0, which is very poorly constrained in most fits. This shows that there is a very strong degeneracy between the broad feature to the blue of the Fe edge at $\sim$ 7 keV in the reflection component \cite{magdziarz1995} and any broad base to the Fe line at 6.4 keV. 

 The EW values obtained from the fits with $E_{fix}$-$\sigma_{free}$ generally are in agreement with those found in previous works based on averaging methods, e.g. \cite{corral2008}, being of 107$\pm$50 eV in the full sample.
 A minimum of 36 \% of this value is in principle the contribution from lines other than the K$_{\alpha}$ neutral iron line, such as the Fe K$_{\beta}$, the Ni K$_{\alpha}$, the Compton Shoulder of the Fe K$_{\alpha}$ \citep{matt2002}, and that is expected to contribute to the detected line. The EW of 107$\pm$50 eV estimated for the full sample would mean that the contribution of the K$_{\alpha}$ iron line only is EW $\sim$69 eV. We expect EW encompassing a range of values from about 50 eV to 100 eV for non-obscured AGN, including the measurements presented here.

Investigating subsamples in luminosity and redshift, we found a hint of a difference in the line EW between the Seyferts subsample and the QSOs one: the line EW, {for example in the second line of each sample in Table \ref{tresults},} is 88$\pm$32 eV in QSOs and 292$\pm$102 in Seyferts. Thus, the EW decreases with the increasing luminosity of the X-ray continuum in agreement with previous results for narrow \citep{iwasawa94} and broad lines \citep{jimenez}. A similar result is found also from the model-independent estimates of the EW presented in the third column of Table \ref{teqw}.

In most cases, if the line width is left free to vary in the spectral fitting, its best fit value becomes higher than zero, as discussed above; on the contrary, if the centroid energy is left free, its output value does not differ from the input 6.4 keV energy (see second and third lines of Table \ref{tresults}).
This indicates that the majority of the line flux is emitted for fluorescence from neutral iron. 

The EW values (e.g. focusing on the $E_{fixed}-\sigma_{fixed}$ fitting presented in Col 9 of Table \ref{tresults}) are consistent with the ones calculated making use of the simulations, presented in Table \ref{teqw}, and explained in Sect. \ref{model_ind}: this confirms that our estimates are robust.

The EW measured in this work are broadly consistent with the values of individual bright sources in the literature: this confirms that the iron lines are emitted by the majority of the sources in our sample because the EW of the minority Iron-line-emitting sources would be much higher than that observed in individual sources. Our resulting EW are also consistent with previous estimates made in iron line stacking, e.g. in \cite{corral2008}.

\subsection{Discussion based on physical models}\label{discussion_physical}
We used a model more specific than a simple Gaussian to fit our data and calculated the EW considering a monochromatic continuum. The 'physical' model is composed by a narrow core centred at 6.4 keV and a relativistic component (\texttt{diskline} in \texttt{xspec}). 

This two-component model is expected in AGN, where reflection of the primary emission is expected to occur both in axis-symmetric regions far away from the BH (such as the torus) and close enough to it for the detection of any line broadening, e.g. in the accretion disk \citep{fabian1989}. 
Both components are in principle expected. 
{Our values of the EW reported in Table \ref{tfix_fix_laor} support this, being consistent with previous results based on averaging techniques and on the study of individual higher quality spectra of AGN in the local Universe. Moreover, our model-independent estimates between 173 and 236 eV (full sample, bulk range in Table \ref{teqw}) are inconsistent with having only a narrow component, which does not commonly exceed 100 eV in the local Universe. This provides an interesting indication, independent of the spectral fitting, of the presence of a relativistic line component in the XMM CDFS data. }
 However, the improvement of the fit by including the relativistic component (Col. 3 in Table \ref{tfix_fix_laor}) is not statistically significant ({e.g., for the full sample, 2.1 $\sigma$ confidence level}), so we have not formally detected a broad component our samples. Having said that, the two line components are strongly correlated and we cannot really disentangle them: we can see the 1, 2, 3 $\sigma$ contours (for two interesting parameters) for the full sample in the space formed by the {EW} of the relativistic line and that of the narrow line in Fig. \ref{contour_abs} {(top panel)}. There are equally-good fits including only each of the components or a combination of both for most samples (see cols. 1, 2 and 9 of Table  \ref{tfix_fix_laor}).
%This degeneracy also explains why the formal 90\% confidence levels of the EW in Table \ref{tfix_fix_laor} {(col. 7 and 8)} almost reach zero eV for both lines, since the $>1\sigma$ $\Delta\chi^2$ contours include those values for either component. 

It is important to notice that our contours (shown for the full sample) exclude the point with zero EW simultaneously for both components at much more than three $\sigma$ confidence level. This indicates a very secure detection of an iron line feature. 
 
 %% {The values of EW=$87\pm83$ eV for the broad component and EW=$48\pm41$ eV for the narrow component obtained in the 'two component model' fitting are consistent with those found in \cite{chaudhary2012}. }
%% {Given that we obtain, for $R$, an upper limit consistent with $R$=1 in most cases, we repeated the fit with this model for the full sample fixing the $R$ parameter to one and found, for the relativistic and narrow component respectively, EW=$88 \pm 87$ eV and EW = $47 \pm 45 $ eV, finding values of the continuum parameters consistent with the ones in Table \ref{tfix_fix_disk}.}

For the full sample, the EW of the narrow line with no relativistic component is $95\pm22$~eV (see first row of Table \ref{tresults}) while the EW of the relativistic line fixing the flux of the narrow line to zero is $347\pm105$~eV (Col. 10 in Table \ref{tfix_fix_laor}, first row). {These results are, within the uncertainties, compatible with} the measurements and upper limits found in \cite{corral2008} and \cite{chaudhary2012}. {Taking into account that our values include contributions of fluorescence in material close to the central SMBH and in outer regions that we are not able to disentangle, they are also consistent, considering the error bars, with the "average" EW=76-143~eV of the relativistic lines detected in individual sources in} \cite{delacalle2010}.
\subsection{Discussion based on comprehensive models}\label{discussion_comprehensive}
\begin{figure}
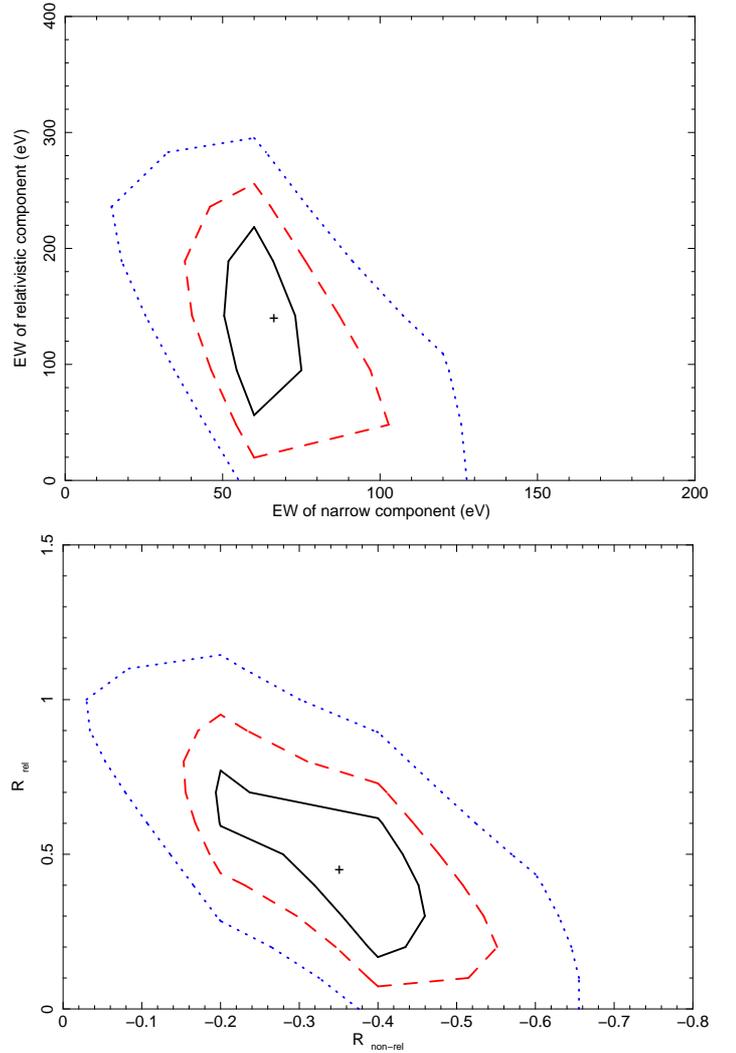

\centering
\includegraphics[width=7cm,angle=-90]{figures/laor_contour_rfix_final.ps}
\includegraphics[width=7cm,angle=-90]{figures/gsmo_pexm_pha_kdblur2_fix_pexm_all_cont_out.ps}
\caption{{Contours at 1 (continuous line), 2 (dashed line), 3 (dotted line) sigma of the EW of the narrow and the relativistic line in the 'two-component model' (top panel). Same contours for the reflection factor $R$ of both the non-relativistic and the relativistic components of the comprehensive model  (bottom panel).} \label{contour_abs} }
  \end{figure}

{We fitted the average spectra with a more complete model with two reflection components. The first one is emitted from the innermost regions of AGN and smeared by relativistic effects due to the vicinity of the central SMBH, the second component is instead emitted from outer regions and is un-blurred.
 The reflection factor contours at 1, 2, 3 sigma, are shown in Fig. \ref{contour_abs}, bottom panel. The figure shows that the relativistic component is different from zero at two sigma confidence level, and the non-relativistic one at three sigma.The two components are coupled: like the above described 'two component' model of the iron line, these cannot be accurately disentangled.
Anyway, we should note that the point with zero reflection factor for both components is excluded at much more than three sigma confidence level, indicating a secure, highly significant detection of reflection features.
}
%Our EW estimate of the narrow line core, of $48\pm41$ eV, is agreement with the values expected from the standard model. 
%The EW value of $87\pm83$ eV measured for the relativistic component in this work is consistent with the upper limit of 300 eV found in \cite{corral2008}.

%% Generally, although our EW of the narrow cores are determined with upper limits in most cases, our measurements are compatible with the values of single sources in the literature. 

%% They are also broadly consistent with previous constrains of average spectra, for example with the EW of 90 eV measured in the average iron line of the XMS-XWAS sample \citep{corral2008}. 
%% The EW of our tentative relativistic components are also consistent with previous works, being of 83 eV in the full sample and ranging from 10 to 168 eV in its subsample: such values lay in the interval of EW found in \cite{delacalle2010}. They are also broadly consistent with the upper limit of 300 eV found in the average spectra of \cite{corral2008}.

\section{Conclusions}\label{conclusions}

We studied a representative sample of 54 AGN (100 spectra from \textsl{XMM-Newton}) with the best spectral S/N in the ultra-deep observation of the \textsl{Chandra} Deep Field South. 
%The long exposure of 3.3 Ms allowed to collect a good number of high quality spectra, with S/N$>15$, totalling more than 180000 counts between 2 and 12 keV rest-frame. 
We investigated several subsamples: two subsamples of column density of the intrinsic absorber (borderline at $ \log{\left( N_{\rm H}\right)} =21.5$), two subsamples in luminosity (with cut at $ \log{\left( L\right)} =44$), and finally two in redshift (segregated at $z=1$). 

We averaged the spectra of the full sample and subsamples developing further our own method previously used in \cite{corral2008} and \cite{falocco2012} to characterise the instrumental and averaging effects. To do this we made use of comprehensive simulations.

 With our new methodology and best data quality so far we could find convincing and solid evidence, using X-ray stacking, for a highly significant iron line in distant AGN, taking into full account the instrumental and methodological effects.
In the spectral fits we convolved our models with a gaussian smoothing with a variable $\Sigma(E)$ introduced by the limited spectral
resolution of the X-ray instruments and the averaging process.

We can summarise our conclusions as follows:
\begin{itemize}

\item The iron feature is significantly detected in the full sample and in all its subsamples.

\item The iron line, modelled with an unresolved Gaussian at 6.4~keV (as expected from neutral material far from the central BH), is detected at 6.8$\sigma$ in the full sample, with EW=$95\pm22$~eV. In the subsamples, the EW measurement ranges from {77} eV for Abs (our objects with $ \log{\left( N_H\right)} \geq21.5$) and 167 eV for the sources with $z<1$ respectively. We find evidence for the Iwasawa-Taniguchi effect comparing the EW of the low luminosity (Seyfert) and high luminosity (QSO) samples. Allowing for a resolved Gaussian profile improves the fit slightly, but not significantly ($\la2\sigma$). 

\item Using a combination of an unresolved line from neutral iron as above and a relativistic line, to take reflection in material far and close to the central BH into account, we found again that the broad component, despite improving the fit moderately in most cases, is not formally required. However, we also found that the fitted parameters of both components are strongly coupled, finding good fits with either or both components. We cannot really disentangle both contributions with our current data. Using a single relativistic component with no unresolved line, we find EW$=347\pm105$~eV for the full sample.

\item The EW of the narrow iron line and that of our tentative relativistic line are both broadly consistent with those (including the upper limits) estimated in previous works based on averaging and stacking methods \citep{corral2008,chaudhary2012} and in individual bright sources in the literature \citep{delacalle2010}, which leads to the conclusion that an iron feature must be widespread among spectra of AGN.

\item The model-independent EW estimates made using our
  thorough simulations produce similar values as those from model
  fitting described above, confirming the robustness of the results
  presented in this paper. Interestingly, the EW {is inconsistent with having a narrow line only and} becomes higher when a broader energy range around the line centroid is considered.{ This suggests} the presence
  of a broad iron feature.

\end{itemize}

Concluding, although the observed lines are well modelled as
fluorescent lines at 6.4 keV in material far away from the primary
emission, we found also some tentative evidence for a component
emitted relatively close to the SMBH. Our work has confirmed that high
S/N and a large amount of spectral counts are required to characterise
average iron features in distant AGN. For a more detailed
characterisation of its physical origin, it is necessary to accumulate
a higher number of spectral counts in the average spectrum.

Although the complexity of the X-ray AGN spectra often makes
challenging the determination of the properties of iron features, with
the deepest \textsl{XMM-Newton} data available and a careful
characterisation of methodological effects, we have confirmed that
iron emission lines represent a universal feature of AGN in the
distant Universe.

\begin{table*}[h] 
  \caption{Results of fits of the average spectrum of the full sample and its subsamples with gsmooth$\otimes$(pha$\times$pex+gauss).\label{tresults}}
  %\centering
 
\scalebox{0.85}{ 
  \begin{tabular}{lrrlllllllllll}
    Sample     &    $\chi^2$/$\nu$(c) & $\chi^2$/$\nu$(g) &$1-P(\Delta\nu,\Delta\chi^2)$  & $\mathrm{N_{\rm H}}$ & $\Gamma$ & $R$   & E  & $\sigma$ & EW  & $\langle z \rangle$ & $\langle \log{\left( L\right)} \rangle$    \\ 
    &  &   &   & $10^{22}$cm$^{-2}$ & - & &  keV  & eV & eV  &   &   erg s$^{-1}$ \\%& $10^{22}cm^{-2}$    \\ 
    (1)    &  (2)   &  (3)  &    (4)  &   (5)  &   (6)   &  (7)  &  (8)   &  (9)  &  (10)  &  (11) &  (12)  \\%& (13) \\ 
    \hline

Full  & 133.4/95 & 87.0/94   &  $9\times10^{-12}$ ($6.8\sigma$)
  & $0.36^{+0.17}_{-0.27}$  & $1.63^{+0.03}_{-0.09}$  & $1.00^{+4.50}_{-0.65}$  &   $\equiv$6.4  & $\equiv$0  & 95$\pm22$  & 1.34 & 43.74   \\%&   1.48\\ 

     &  & 86.2/93   & [$0.4$ ($0.9\sigma$)]
  & $0.36\pm0.22$  &  $1.63^{+0.04}_{-0.08}$ & $1.0^{+3.8}_{-1.0} $ &    $\equiv$6.4  & $92^{+236}_{-92}$  & 107$\pm50$  &   &  \\%&    \\ 

     &  &  87.0/93  & [$1$ ($0\sigma$)]
  & $0.36\pm0.21$  &  $1.63\pm0.06$ & $1.0^{+4.1}_{-0.6}$ &    6.4$\pm$0.20  & $\equiv0$  & 94$\pm22$  &   & \\% &    \\ 

     &  &  85.9/92  & [$0.6$ ($0.52\sigma$)]
  & $0.36\pm0.21$  &  $1.63\pm0.05$ & $1.0^{+3.7}_{-1.0} $ &    6.44$_{-0.10}^{+0.04}$  & $91_{-87}^{+197}$  & $109\pm44$  &   &  \\%&    \\ 

   \hline

%% Unabs   &  124.8/95  &  92.0/94  & $1\times10^{-8}$ ($5.7\sigma$) & %  $10^{-8}
%% $<0.82$ & $1.82\pm0.03$ & $1.0^{+4.2}_{-0.4}$ &  $\equiv$6.4  & $\equiv$0  & 101$\pm25$  & 1.37  & 43.78 \\% &   0.05  \\ 

%%  &    &  91.9/93    & [$0.7$ ($0.4\sigma$)]
%%  & %  $10^{-8}
%% $<0.82$ & $1.82\pm0.03$ & $1.0^{+0.7}_{-0.4}$ &  $\equiv$6.4  &  $50^{+100}_{-50}$ & 106$\pm41$  &   &  \\% &    \\ 

%%  &    &   92.0/93   & [$1$ ($0\sigma$)]
%%  &  % $10^{-8}
%% $<0.82$ & $1.82\pm0.03$ & $1.0^{+0.7}_{-0.4}$ &  6.40$\pm$0.05  &  $\equiv$0 & 101$\pm25$  &   & \\%  &    \\ 
%%  &    &   91.9/92   & [$0.1$ ($0.1\sigma$)]
%%  & %  $10^{-8}
%% $<0.82$ & $1.82\pm0.03$ & $1.0^{+0.7}_{-0.4} $ &  6.40$_{-0.13}^{+0.28}$  &  $51_{-50}^{+820}$ & $106\pm33$  &   &  \\% &    \\ 

{Unabs}   &  139.2/95  &  108.7/94  & $3\times10^{-8}$ ($5.5\sigma$) & %  $10^{-8}
$<0.82$ & $1.87\pm0.03$ & $1.0\pm0.4$ &  $\equiv$6.4  & $\equiv$0  & $107\pm32$  &  1.27 & 43.78 \\% &   0.05  \\ 

 &    &  108.7/93    & [$1$ ($0 \sigma$)]
 & %  $10^{-8}
$<0.82$ & $1.87\pm0.03$ & $1.0^{+0.2}_{-0.2}$ &  $\equiv$6.4  &  $1^{+822}_{-1}$ & $107\pm32$  &   &  \\% &    \\ 

 &    &   108.7/93   & [$1$ ($0\sigma$)]
 &  % $10^{-8}
$<0.82$ & $1.87\pm0.03$ & $1.0^{+0.2}_{-0.2}$ &  $6.4_{-0.1}^{-0.4}$  &  $\equiv$0 & $107\pm32$  &   & \\%  &    \\ 
 &    &   108.7/92   & [$1$ ($0\sigma$)]
 & %  $10^{-8}
$<0.82$ & $1.88\pm$0.03 & $1^{+0.2}_{-0.2} $ &  $6.4_{-0.1}^{+0.3}$  &  $5_{-5}^{+800}$ & $107\pm32$  &   &  \\% &    \\ 

   \hline

%Abs    &  110./95  & 95.9/94  & $1.7\times10^{-4}$ ($3.8\sigma$) &   $2.95\pm0.50$ & $1.52\pm0.08 $  & $1.0^{+5.5}_{-1}$ &  $\equiv$6.4  & $\equiv$0  & 75$\pm35$  & 1.34  & 43.74 \\% &  4.51  \\ 

{Abs }   &  109.4/95  & 94.2/94  & $1\times10^{-4}$ ($3.9\sigma$) &   $3.67\pm$0.55 & $1.49\pm0.08 $  & $1^{+3.6}_{-1}$ &  $\equiv$6.4  & $\equiv$0  & $77\pm34$  &  1.28 & 43.74 \\% &  4.51  \\ 

% &   &  82.4/93    &  [$2.4\times10^{-4}$ ($3.7\sigma$)]  &  $2.64\pm0.50$ & $1.43\pm0.08$ & $0.001^{1.800}_{-0.001}$ &  $\equiv$6.4  & $660^{+340}_{-262}$  & 336$\pm180$  &   & \\%  &    \\ 

 &   &  85.4/93    &  [$3\times10^{-3}$ ($3\sigma$)]  &  $3.39\pm$0.53 & $1.39\pm$0.08 & $0.003^{+1.100}_{-0.001}$ &  $\equiv$6.4  & $455^{+294}_{-187}$  & $243\pm123$  &   & \\%  &    \\ 

% &   &   95.9/93    &  [$1$ ($0\sigma$)]&  $2.96\pm0.50$ & $1.52\pm0.08$ & $1.0^{+5.5}_{-1.0}$ &  6.40$\pm$0.25  & $\equiv$0  & 75$\pm35$  &   &  \\% &    \\ 

&   &   94.2/93    &  [$1$ ($0\sigma$)]&  $3.67\pm0.55$ & $1.52^{+0.10}_{-0.09}$ & $1^{+3.7}_{-1}$ &  $6.4\pm0.2$  & $\equiv$0  & $77\pm32$  &   &  \\% &    \\ 

% &   &  82.1/92    &  [$0.003$ ($3.3\sigma$)]  &  $2.60\pm0.50$ & $1.43\pm0.07$ & $0.003^{+1.500}_{-0.003}$ &  6.35$\pm0.30$  & $700^{+300}_{-250}$  & $348\pm177$  &   &  \\% &    \\ 

&   &  85.6/92    &  [$3\times10^{-3}$ ($2.9\sigma$)]  &  $3.42\pm0.54$ & $1.39\pm0.08$ & $0.003^{+1.050}_{-0.003}$ &  $6.47\pm0.20$  & $438^{+273}_{-192}$  & $239\pm109$  &   &  \\% &    \\ 

   \hline

 z$<$1  &  112.6/95   &  97.0/94  & $8\times10^{-5}$ ($3.9\sigma$) 
&   $0.86\pm0.40$ & $1.58\pm 0.08$  & $1.0^{+13.5}_{-1.0} $ &  $\equiv$6.4  & $\equiv$0  & 167$\pm$67  & 0.64  & 43.21 \\% &  1.61  \\ 
  &    &  94.0/93   & [$0.08$ ($1.7\sigma$)]
  &   $0.92\pm0.40$ & $1.56\pm0.08$ & $1.0^{+7.5}_{-0.6}$ &  $\equiv$6.4  &  $4^{+150}_{-4}$ & 81$\pm34$  &   &  \\% &    \\ 

  &    &  94.0/93   & [$0.08$ ($1.7\sigma$)]
  &   $0.92\pm0.40$ & $1.56\pm0.08$ & $1.0^{+7.5}_{-0.6}$ &  6.4$^{+0.6}_{-0.4}$  &  $\equiv$0 & 81$\pm34$  &   &  \\% &    \\ 
  &    &  94.0/92   & [$0.22$ ($1.2\sigma$)]
  &   $0.92\pm0.40$ & $1.56\pm0.08$ & $1.0^{+0.9}_{-0.6}$ &  6.40$\pm_{-0.13}^{+0.28}$  &  $2_{-2}^{+822}$ & $81\pm34$  &   &  \\% &    \\ 

   \hline

  z$\geq$1  &  109.9/95  &   78.5/94  & $2\times10^{-8}$ ($5.6\sigma$) 
&   %$0.121\pm0.121$
$<0.24$ & $1.63\pm0.06$  & $0.3^{+1.6}_{-0.3}$ &  $\equiv$6.4  & $\equiv$0  & 114$\pm30$  & 1.80  & 44.07  \\%&  1.39   \\ 
 &    &   76.2/94   & [$0.1$ ($1.5\sigma$)]
 &  % $0.07^{+0.25}_{-0.07}$
$<0.32$ & $1.60\pm0.05$ & $0.008^{+1.300}_{-0.008}$ & $\equiv$6.4  &  $150^{+100}_{-152}$ & 114$\pm42$  &   &   \\%&    \\ 
 &    &   78.5/93   & [$1$ (0$\sigma$)]
 & %  $0.121^{0.320}_{0.121}$
$<0.441$ & $1.63^{+0.11}_{-0.04}$ & $0.3^{+1.7}_{-0.3}$ & 6.40$^{+0.25}_{-0.25}$  &  $\equiv$0 & 150$\pm40$  &   &  \\% &    \\ 
 &    &   76.2/92   & [$0.32$ ($1\sigma$)]
 &   %$0.07^{+0.25}_{-0.07}$
$<0.32$ & $1.60^{+0.05}_{-0.04}$ & $0.008_{-0.008}^{+1.400}$ & 6.40$^{+0.10}_{-0.07}$  &  $151^{+117}_{-151}$ & $150\pm55$  &   &  \\% &    \\ 

   \hline

%\scalebox{0.90}{$\log{\left( L/{\rm erg~s^{-1}}\right)}<$44}

  Sey &   129.9/95 &  93.9/94   & $2\times10^{-9}$ ($6\sigma$) 
&   $0.36^{-0.30}_{-0.29}$ & $1.58\pm0.07  $  & $1.0^{+2.5}_{-0.8}$ &  $\equiv$6.4  & $\equiv$0  & 103$\pm27$  & 0.91  & 43.35 \\% & 1.35   \\ 
 &    &   90.6/93   & [$0.066$ ($1.84\sigma$)]
 &   %$0.235_{-0.235}^{+0.235}$
$<0.47$ & $1.51\pm0.07$ & $0.004_{-0.004}^{+1.800}$ & $\equiv$6.4  &  $450^{+200}_{-200}$ & 293$\pm102$  &   &  \\% &    \\ 
  &   & 93.9/93     & [$1$ ($0\sigma$)] 
&   $0.36^{+0.29}_{-0.29}$ & $1.58\pm0.07$  & $1.0_{-0.8}^{+2.5} $ &  6.40$_{-0.05}^{+0.05}$  & $\equiv$0  & 103$\pm27$  &    &  \\% &   \\ 
 &   & 90.4/92 & [$0.17$ ($1.4\sigma$)] & %  $0.254^{+0.308}_{-0.254}$
$<0.56$ & $1.51\pm0.06$  & $0.004_{-0.004}^{+1.800} $ &  $6.5_{-0.20}^{+0.09}$  & 425$\pm187$  & 289$\pm96$  &    &   \\%&   \\ 
   \hline

%\scalebox{0.90}{$\log{\left( L/{\rm erg~s^{-1} }\right)}\geq$44}
 QSO &  204.2/95 &  178.2/94   & $3\times10^{-7}$ ($5.10\sigma$) 
&   $0.11^{+0.23}_{-0.08}$ & $1.60_{-0.02}^{+0.07}$  & $0.14_{-0.14}^{+1.36} $ &  $\equiv$6.4  & $\equiv$0  & 119$\pm30$  &  2.02  &  44.34 \\%&  1.68  \\ 

 &    &  166.6/93   & [$5.3\times10^{-4}$ ($3.5\sigma$)]
 &   $0.20\pm0.15$ & $1.65\pm0.04$ & $0.66^{+1.34}_{-0.04}$ &  $\equiv$6.4  &  $101^{+120}_{-100}$ & 88$\pm32$  &   &   \\%&    \\ 
 &    & 167.3/93    & [$9\times10^{-4}$ ($3.3\sigma$)]
 &   $0.24\pm0.15$ & $1.68\pm0.4$ & $0.9^{+1.3}_{-0.8}$ &  6.40$_{-0.15}^{+0.05}$  &  $\equiv$0 & 75$\pm22$  &   &   \\%&    \\ 

 &    &  164.9/92   & [$0.0013$ ($3.22\sigma$)]
 &   $0.19\pm0.15$ & $1.65\pm0.04$ & $0.61^{+0.30}_{-0.30}$ &  6.34$_{-0.07}^{+0.07}$  &  $130^{+89}_{-112}$ & $93\pm31$  &   &  \\% &    \\ 

  \end{tabular}
  %% for referee version 
 }
  \tablefoot{Columns: (1) sample; (2) $\chi^2$/dof of the fit with the continuum; (3) $\chi^2$/dof of the fit with the continuum and the Gaussian; (4) probability $P(\Delta \chi^2, \Delta \nu) $ (of adding the Gaussian to the model and of leaving each parameter free to vary, see text); (5) intrinsic column density estimated in the fits; (6) slope of the powerlaw; (7) reflection factor; (8) central energy of the Gaussian; (9) $\sigma$ of the Gaussian; (10) EW of the Gaussian; columns 11 (average redshift of the sample) and 12 (average logarithmic luminosity of the sample in erg s$^{-1}$), are repeated from Table 1 for reference }
\end{table*}

\begin{table*}[h] 
  \caption{{Fits results of the average spectrum of the full sample and subsamples using} gsmooth$\otimes[$pha$\times$(pex+laor)+gaus)$]$.\label{tfix_fix_laor}}
  %\centering
  %for referee version: 
  \scalebox{0.95}{ 
  \begin{tabular}{lrrllllll|rl}
    Sample    
& $\chi^2_{g}$/$\nu$ 
    & $\chi^2$/$\nu$ 
    & $1-P(\Delta\nu,\Delta\chi^2)$
%& S
    & $\mathrm{N_{\rm H}}$ 
    & $\Gamma$ 
    & $Rel_{refl}$  
    %% & $R_i$    
    %% & $\beta$   
    %% &   $\theta$ 
    
    &    EW$_{rel}$  
    & EW$_{na}$ 
& $\chi^2_{one-rel}$/$\nu$ 
 &    EW$_{one-rel}$  
 \\
 
 &   &  &    &10$^{22}$cm$^{-2}$ &  &  &eV & eV  &   & eV \\ 
 (0) &   (1) & (2) & (3) & (4) & (5) & (6) & (7) & (8) & (9) & (10)  \\
    \hline
    Full  &  87.0/94  & 82.6/93  &   3.6$\times10^{-2}$($2.1\sigma$)  & $0.24\pm 0.24$ &   $1.61\pm0.06 $ &  $1.0_{-1.0}^{+2.3}$ %& $\equiv$6 & $\equiv$-2  & $\equiv$30   
&  $140\pm120$ &  $67\pm28$ & 98.4/94 & $347\pm105$  \\ 

  \hline

  Unabs   &  108.7/94  & 108.7/93  &   1($0  \sigma$)  &%10^{-8} 
$<0.82 $ &   $1.87\pm0.03$ &  $1_{-0.2}^{+0.4}$   &  $0_{-0}^{+1000}$ &  $107_{-32}^{+32}$  & 138.9/94 & $0_{-0}^{+1000}$ \\ 

   \hline
 Abs   & 94.2/94 &  92.6/93 &   0.2($1.3\sigma$)  & $2.84\pm0.60 $ &   $1.29 \pm0.08 $ &  $0. _{-0}^{+2}$    &  $404\pm140$ & %1
 $<1000$ & 92.6/94 & $404\pm140$  \\

 \hline

z$<$1  &  97.0/94  & 95.2/93  &  0.2($1.3\sigma$)   & $0.68\pm0.40 $ &   $  1.55\pm 0.09 $ &  $1_{-1}^{+4}$   &  $173\pm173$ & %1
 $104\pm93$ & 99.7/94 & $362\pm210$  \\ 
   \hline

 z$\geq$1  & 78.5/94    &  77.6/93 &  0.3($1\sigma$)   & %$0.06^{+0.25}_{-0.06} $
$<0.3$ &   $1.58 \pm 0.04$ &  $0.007_{-0.007}^{+1.680}$   &  $84\pm84$ &  $98_{-31}^{+31}$ & 94.6/94 & $318\pm119$   \\

   \hline

 Sey    & 93.9/94   & 92.2/93  
&  0.2($1.3\sigma$)   & $0.26 ^{+0.32}_{-0.26} $ &   $1.57\pm 0.07$ &  $1_{-1}^{+6} $   &  $118\pm118$ &  $82_{-35}^{+35}$  &  107.6/94  & $380\pm130$  \\

   \hline

 QSO   & 178.2/94  &  166.7/93 
&  7$\times10^{-4}$($3.4\sigma$)   &% $0.1\pm0.1 $ 
$<0.138$ &   $1.58\pm0.08 $ &  $0.13_{-0.13}^{+0.80} $   &  $228\pm70$ &  $49\pm45$ & 169.7/94 &  $316\pm82$   \\

 \hline  \hline

  \end{tabular}
  %% for referee version }
}
  \tablefoot{{In all fits we have fixed  R$_i\equiv$6, $\beta\equiv$-2, $\theta\equiv$30. Columns: (0) sample; (1) $\chi^2$/dof of the fit with the Gaussian with $E_{fixed}-\sigma_{fixed}$ only, repeated from Table \ref{tresults} for reference; (2) $\chi^2$/dof with the two-component model (see text); (3) Significance of the \texttt{laor}  (expressed in 1-$P(\Delta\nu,\Delta\chi^2)$ and in $\sigma$); (4) intrinsic column density of the neutral absorber. (5) powerlaw slope. (6) reflection factor; (7) \texttt{laor} EW measured from the two-component model; (8) EW of the narrow component; (9) $\chi^2$/dof of the fit after having fixed the flux of the narrow line component to zero; (10) \texttt{laor} EW measured from the same fit of Col. 9. }}
 \end{table*}

\begin{table*}[h] 
  \caption{{Fits results of the average spectrum of the full sample and subsamples using \texttt{gsmooth$\otimes[$ pha $\times$ kdblur2 $\otimes$(pexm)+pexm)$]$} }.  \label{tpexm}}
  %\centering
  %for referee version: 
  \begin{tabular}{lrllll}
    Sample   
    & $\chi^2$/$\nu$ 
    & $\mathrm{N_{\rm H}}$ 
    & $\Gamma$ 
    & $R_{non-rel}$  
& $R_{rel}$  
 \\
  (0) & (1) & (2) & (3) & (4) & (5)   \\
    \hline
    Full  & 84.1/94  &  0.46$\pm 0.03$  & 1.71$\pm$0.06 &   -0.35$\pm $0.18 &  0.45$_{-0.35}^{+0.38}$  \\ 
   Unabs  & 104.0/94  &  $<$0.10   & 1.95$\pm$0.06 &   -0.50$\pm$0.27 &  0.43$_{-0.43}^{+0.54}$  \\ 
   Abs  & 90.2/94  &  3.68 $\pm $0.53  & 1.56$\pm $0.09 &   -0.167$_{-0.228}^{+0.168} $ &  $0.59_{-0.46}^{+0.51}$  \\ 
   z$<1$  &  91.5/94 &  1.01$\pm$0.40   & 1.65$\pm$0.10 &   -0.317$\pm$0.220 &  $0.46_{-0.43}^{+0.49}$  \\
   z$\geq1$     & 80.5/94  &  0.343$\pm$0.290   & 1.76$\pm$0.07 &   -0.53$\pm $0.23 &  $0.08_{-0.08}^{+0.48}$  \\ 
   Sey     & 89.0/94  &  0.48$\pm$0.30   & 1.69$\pm$0.09 &   -0.39$\pm$0.22 &  $0.54\pm0.45$  \\ 
  QSO    & 73.9/94  &  0.57$\pm$0.32   & 1.76$\pm $0.08 &   -0.33$\pm$0.26 &  0.16$_{-0.16}^{+0.49}$  \\ 
  \end{tabular}
  %% for referee version }
  \tablefoot{ {In the fits we have fixed: $R_{in}$=1.235$R_g$, $R_{br}$=20$R_g$, $R_{out}$=400$R_g$. Columns: (0): sample; (1): $\chi^2$/$\nu$ of the fit; (2): intrinsic column density in 10$^{22}cm^{-2}$; (3): photon index; (4):  reflection factor of the reflection component produced far away from the central engine; (5): reflection factor of the reflection component produced close to the central engine. }}
 \end{table*}

%% \begin{table*}[h] 
%%   \caption{Fits results of the average spectrum of the full sample and subsamples using \texttt{gsmooth$\otimes[$ pha $\times$ kdblur2 $\otimes$(pexm)+pexm)$]$}.\label{tpexm}}
%%   %\centering
%%   %for referee version: 
  
%%   \begin{tabular}{lllll}
%%   (1)   & (2) & (3)& (4)& (5) \\
 
%% $\chi^2$/dof & 471/94 & 190/93 & 228/93 & 109/92 \\
%% $\Gamma$ & 2.5 & 1.26 &  1.65 $^{+0.60}_{-0.16}$ & 1.76$^{+0.02}_{-0.17}$ \\
%% $R_1$ &-0.007$^{+0.787}_{-0.001}$ &-0.05 &- 0.14$^{+0.02}_{-0.04}$ & -0.49$^{+0.07}_{-0.20}$ \\
%% A & $\equiv$1 & $\equiv$1  & 0.27$\pm$ & 0.17$^{+0.06}_{-0.02}$ \\
%% $N_H$  & 2.64$\pm0.12$ &2.59 &2.59$\pm0.09$ &1.50$\pm0.09$ \\
%% $R_b$   &- & &- &1.23$_{-1.23}^{+0.15}$ \\
%% $\Gamma$  &2.5 & 2.5 & 2.5 & 2.5 \\
%% $R_2$ & 3.5$_{-6}^{0.2}$ & 5.43$ $ & $10^{-7}$ & 0.24$^{+0.55}_{-0.12}$ \\

%%  \end{tabular}
%%  \end{table*}

%% END TABLE %%%%%%%%%%%%%%%%%%%%%%%%%%%%%%%%%%%%%%%%%%%%%%%%%%%%%%%%%%%%%%%%%%%

\begin{acknowledgements}
      Financial support for this work was provided by the Spanish Ministry of Economy and Competitiveness through the grant \emph{AYA2010-21490-C02-01}.\\
      The authors acknowledge the team of the Spanish Supercomputing Network (RES) node (Altamira) at the Universidad de
Cantabria in Santander whose facilities helped to improve the computing time of this work. \\
We acknowledge financial contribution from the agreement ASI-INAF I/009/10/0 and from the PRIN-INAF 2011.\\
PR acknowledges a grant from the Greek General Secretariat of Research and Technology in the
framework of the programme Support of Postdoctoral Researchers.

\end{acknowledgements}

\bibliographystyle{aa}
\bibliography{bibtex}

\end{document}